\documentclass[a4paper,fleqn,usenatbib]
{mnras}
\usepackage{newtxtext,newtxmath}
\usepackage[T1]{fontenc}
\usepackage{ae,aecompl}

\usepackage{graphicx}	
\usepackage{amsmath}	
\usepackage{amssymb}	
\usepackage{epsfig}
\usepackage{pdflscape}
\usepackage{natbib}
\newcommand{\ergs}{${\rm erg \ cm^{-2} \ s^{-1}}$}
\newcommand{\erg}{${\rm erg \ s^{-1}}$}
\newcommand{\arcsinh}{{\rm arcsinh} }

\newcommand{\todo}{\ifmmode {\Huge \bullet} \else {\Huge$\bullet$}\fi}

\newcommand{\vFWHM}{\ifmmode V_{\mbox{\tiny FWHM}} \else $V_{\mbox{\tiny FWHM}}$ \fi}

\newcommand{\kms}{\ifmmode {\rm km\,s}^{-1} \else km\,s$^{-1}$ \fi}

\newcommand{\ergcms}{\ifmmode {\rm ergs\,cm}^{-2}\,{\rm s}^{-1} \else ergs\,cm$^{-2}$\,s$^{-1}$\fi}
\newcommand{\ergcmsA}{\ifmmode{\rm ergs}\, {\rm cm}^{-2}\,{\rm s}^{-1}\,{\rm\AA}^{-1} \else ergs\, cm$^{-2}$\, s$^{-1}$\, \AA$^{-1}$\fi}
\newcommand{\ergcmsHz}{\ifmmode{\rm ergs\,cm}^{-2}\,{\rm s}^{-1}\,{\rm Hz}^{-1} \else ergs\,cm$^{-2}$\,s$^{-1}$\,Hz$^{-1}$\fi}
\newcommand{\phcms}{\ifmmode {\rm ph\,cm}^{-2}\,{\rm s}^{-1} \else ,ph\,cm$^{-2}$\,s$^{-1}$\fi}
\newcommand{\phcmsA}{\ifmmode {\rm ph\,cm}^{-2}\,{\rm s}^{-1}\,{\rm\AA}^{-1} \else ph\,cm$^{-2}$\,s$^{-1}$\,\AA$^{-1}$\fi}

\newcommand\Msun{\ifmmode M_{\odot} \else $M_{\odot}$\fi}
\newcommand\msun{\ifmmode M_{\odot} \else $M_{\odot}$\fi}
\newcommand\Lsun{\ifmmode L_{\odot} \else $L_{\odot}$\fi}
\newcommand\Zsun{\ifmmode Z_{\odot} \else $Z_{\odot}$\fi}
\newcommand\mpyr{\ifmmode \Msun\,{\rm yr}^{-1} \else $\Msun\,{\rm yr}^{-1}$ \fi}

\newcommand{\Luv}{\ifmmode L_{1450} \else $L_{1450}$\fi}
\newcommand{\Lop}{\ifmmode L_{5100} \else $L_{5100}$\fi}
\newcommand{\Lthree}{\ifmmode L_{3000} \else $L_{3000}$\fi}
\newcommand{\lledd}{\ifmmode L/L_{\rm Edd} \else $L/L_{\rm Edd}$\fi}
\newcommand{\ledd}{\ifmmode L_{\rm Edd} \else $L_{\rm Edd}$\fi}
\newcommand{\lamLlam}{\ifmmode \lambda L_{\lambda} \else $\lambda L_{\lambda}$\fi}
\newcommand{\lbol} {\ifmmode L_{\rm bol} \else $L_{\rm bol}$\fi}
\newcommand{\llbol}{\ifmmode \log\left(\lbol/\ergs\right) \else $\log\left(\lbol/\ergs\right)$\fi}

\newcommand{\fuv}{\ifmmode f_{\lambda}\left(1450\AA\right) \else $f_{\lambda}\left(1450 {\rm \AA}\right)$\fi}
\newcommand{\fthree}{\ifmmode f_{\lambda}\left(3000\AA\right) \else $f_{\lambda}\left(3000{\rm \AA}\right)$\fi}
\newcommand{\fH}{\ifmmode f_{\lambda}\left(1.65\micron\right) \else
$f_{\lambda}\left(1.65\micron\right)$\fi}

\newcommand{\mbh}{\ifmmode M_{\rm BH} \else $M_{\rm BH}$\fi}
\newcommand{\lmbh}{\ifmmode \log\left(\mbh/\Msun\right) \else $\log\left(\mbh/\Msun\right)$\fi}

\newcommand \Hbeta {\ifmmode {\rm H}\beta \else H$\beta$\fi}
\newcommand \hb    {\ifmmode {\rm H}\beta \else H$\beta$\fi}
\newcommand  \mgii  {\ifmmode {\rm Mg}{\textsc{ii}} \else Mg\,{\sc ii}\fi}
\newcommand  \MGII  {\ifmmode {\rm Mg}\,{\sc ii}\,\lambda2798 \else Mg\,{\sc ii}\,$\lambda2798$\fi}
\newcommand  \siiv  {\ifmmode {\rm Si}\, {\sc iv}\ \else Si\,{\sc iv}\fi}
\newcommand  \SIIV  {\ifmmode {\rm Si}\,{\sc iv}\,\lambda1399 \else Si\,{\sc iv}\,$\lambda1399$\fi}
\newcommand  \civ  {\ifmmode {\rm C}\, {\sc iv}\ \else C\,{\sc iv}\fi}
\newcommand  \CIV  {\ifmmode {\rm C}\,{\sc iv}\,\lambda1549 \else C\,{\sc iv}\,$\lambda1549$\fi}
\newcommand  \NV  {\ifmmode {\rm N}\,{\sc v}\,\lambda1240 \else N\,{\sc v}\,$\lambda1240$\fi}
\newcommand  \nv  {\ifmmode {\rm N}\,{\sc v}\ \else N\,{\sc v}\fi}
\newcommand  \LyA  {\ifmmode {\rm Ly}\,{\sc $\alpha$}\,\lambda1216 \else Ly\,{\sc $\alpha$}\,$\lambda1216$\fi}
\newcommand  \lya {\ifmmode {\rm Ly}\,{\sc $\alpha$}\ \else Ly\,{\sc $\alpha$}\fi}
\newcommand  \feii {\ifmmode {\rm Fe}\,{\textsc{ii}}\, \else Fe\,{\sc ii}\fi}

\newcommand  \aliii  {\ifmmode {\rm Al}{\textsc{iii}} \else Al\,{\sc iii}\fi}

\newcommand  \CIII  {\ifmmode {\rm C}\,{\sc iii]}\,\lambda1909 \else C\,{\sc iii]}\,$\lambda1909$\fi}
\newcommand  \oi    {\ifmmode \left[{\rm O}\,{\textsc i}\right] \else [O\,{\sc i}]\fi}
\newcommand  \OI    {\ifmmode \left[{\rm O}\,{\textsc i}\right]\,\lambda6300 \else [O\,{\sc i}]$\,\lambda6300$ \fi}

\newcommand  \oii   {\ifmmode \left[{\rm O}\,{\textsc ii}\right] \else [O\,{\sc ii}]\fi}
\newcommand  \OII   {\ifmmode \left[{\rm O}\,{\textsc ii}\right]\,\lambda3727 \else [O\,{\sc ii}]\,$\lambda3727$ \fi}
\newcommand  \oiii  {\ifmmode \left[{\rm O}\,{\textsc iii}\right] \else [O\,{\sc iii}]\fi}
\newcommand  \OIII  {\ifmmode \left[{\rm O}\,{\textsc iii}\right]\,\lambda5007 \else [O\,{\sc iii}]\,$\lambda5007$\fi}
\newcommand  \loiii  {\ifmmode L_{\left[{\rm O}\,{\textsc iii}\right]} \else $L_{[\rm O\,{\rm III}]}$\fi}

\newcommand{\lmg}{\ifmmode L\left(\mgii\right) \else $L\left(\mgii\right)$\fi}
\newcommand{\fwmg}{\ifmmode {\rm FWHM}\left(\mgii\right) \else FWHM(\mgii)\fi}
\newcommand{\fwciv}{\ifmmode {\rm FWHM}\left(\civ\right) \else FWHM(\civ)\fi}
\newcommand{\fwhm}{\ifmmode {\rm FWHM} \else FWHM\fi}

\date{Accepted XXX. Received YYY; in original form ZZZ}
\pubyear{2018}

\begin{document}
\title[Morphology of AGN Emission Line Regions]{Morphology of AGN Emission Line Regions in SDSS-IV MaNGA Survey} 

\author[He et al.]{
Zhicheng He$^{1,2,3}$\thanks{E-mail: zcho@jhu.edu}, 
Ai-Lei Sun$^{1}$\thanks{E-mail: asun27@jhu.edu}, 
Nadia L. Zakamska$^{1}$\thanks{E-mail: zakamska@jhu.edu}, 
Dominika Wylezalek$^{4}$, 
\newauthor
Michael Kelly$^{5}$, Jenny E. Greene$^{6}$, Sandro B. Rembold$^{7,9}$, 
Rog\'erio Riffel$^{8,9}$ and
\and Rogemar A. Riffel$^{7,9}$
\\
$^{1}$Department of Physics \& Astronomy, Johns Hopkins University, Bloomberg Center, 
3400 N. Charles St., Baltimore, MD 21218, USA\\
$^{2}$CAS Key Laboratory for Research in Galaxies and Cosmology, Department of Astronomy, University of Science and 
Technology of China, \\
Hefei, Anhui 230026, China\\
$^{3}$School of Astronomy and Space Science, University of Science and 
Technology of China, Hefei 230026, China\\
$^{4}$European Southern Observatory, Karl-Schwarzschildstr. 2, 85748 Garching bei M\"{u}nchen, Germany\\
$^{5}$Applied Physics Laboratory, Johns Hopkins University, 11100 Johns Hopkins Rd., Laurel MD 20273, USA\\
$^{6}$Department of Astrophysical Sciences, Princeton University, Princeton, NJ 08544, USA\\
$^{7}$Departamento de F\'\i sica, CCNE, Universidade Federal de Santa Maria, 97105-900, Santa Maria, RS, Brazil\\
$^{8}$Departamento de Astronomia, Universidade Federal do Rio Grande do Sul, IF, CP 15051, 91501-970 Porto Alegre, RS, Brazil\\
$^{9}$Laborat\'orio Interinstitucional de e-Astronomia - LIneA, Rua Gal. Jos\'e Cristino 77, Rio de Janeiro, RJ - 20921-400, Brazil\\
}
\maketitle
\begin{abstract}
Extended narrow-line regions (NLRs) around active galactic nuclei (AGN) are shaped by the distribution of gas in the host galaxy and by the geometry of the circumnuclear obscuration, and thus they can be used to test the AGN unification model.
In this work, we quantify the morphologies of the narrow-line regions in 308 nearby AGNs ($z=0-0.14$, \lbol $\sim 10^{42.4-44.1}$ \erg{}) 
from the MaNGA survey. 
Based on the narrow-line region maps, we find that a large fraction (81\%) of these AGN have bi-conical NLR morphology. The distribution of their measured opening angles suggests that the intrinsic opening angles of the ionization cones has a mean value of 85--98$^\circ$ with a finite spread of 39--44$^\circ$ (1-$\sigma$).
Our inferred opening angle distribution implies a number ratio of type I to type II AGN of 1:1.6--2.3, consistent with other measurements of the type I / type II ratio at low AGN luminosities.
Combining these measurements with the WISE photometry data,
we find that redder mid-IR color (lower effective temperature of dust) corresponds to stronger and narrower photo-ionized bicones. 
This relation is in agreement with the unification model that suggests that the bi-conical narrow-line regions are shaped by a toroidal dusty structure within a few pc from the AGN. 
Furthermore, we find a significant alignment between the minor axis of host galaxy disks and AGN ionization cones.
Together, these findings suggest that obscuration on both circumnuclear ($\sim $pc) and galactic ($\sim$ kpc) scales are important in shaping and orienting the AGN narrow-line regions. 

\end{abstract}


\section{Introduction} \label{sec:intro}

The unification scheme of active galactic nuclei (AGN; \citealt{Antonucci1993}) was developed to resolve the dichonomy 
of AGN types with one geometrical model. 
The model posits that all AGN are fundamentally the same and are powered by a nucleus emitting featureless continuum and broad lines 
which is embedded in a torus-like dusty structure. 
The two spectral types, broad line (type I) and narrow line (type II) AGN, arise depending on whether the observer has a direct view to the nucleus (type I) or whether the dusty torus blocks our line of sight (type II).
The viewing angle is the sole factor in determining the spectral type of the AGN, as the model presumes a universal dusty torus that is uniform, opaque, and has a fixed opening angle at any given AGN luminosity \citep[]{Antonucci1993}. 
This model elegantly explains a number of observables, including the polarized continuum \citep{Antonucci1985}, conical scattered light and narrow-line regions \citep[e.g.,][]{Evans1991,Mulchaey1996,Mulchaey1996a,Zakamska2005,Obied2016}, and dust emission in the mid-IR \citep{Pier1992,Pier1993}. 

Even at a fixed luminosity, AGN differ in properties such as radio brightness or presence of outflows, so it is reasonable to expect 
variations in the dusty torus geometry as well. 
Evidence from optical and mid-IR observations suggests that AGN tend to have a range of torus covering factors \citep[e.g.,][]{RamosAlmeida2011,Muller-Sanchez2011,Ichikawa2015,Audibert2017, Garcia-Gonzalez2017}. 
It is important to quantify the distribution of the torus covering factor, because it can cause biases in AGN selection with respect to spectral types, in the sense that AGN with larger covering factors are more likely to be classified as type II and vise versa \citep{Elitzur2012}. 
However, direct measurements of the distribution of the opening angle has been challenging. 

Near-IR and mid-IR spectroscopic modeling of dust emission can constrain the covering factor to some extent, but has strong degeneracies \citep[][]{Netzer2015}. 
Resolving the spatial structure of the torus has been limited to a handful of nearby systems, and even so their inner edge cannot be seen \citep[e.g.,][]{Hoenig2010,Tristram2014}. 
Currently, the best estimates rely on indirect methods, such as the ratio of hot dust emission to the AGN bolometric luminosity (\lbol), inferred from X-ray, optical, or both, \citep[e.g.,][]{Maiolino2007,Treister2008,Mateos2016,Ezhikode2017}, but such measurements are subject to systemic uncertainties in inferring the total torus and AGN luminosity and depend on assumptions on their emission anisotropy. 

Statistical studies of narrow-line region morphology offer a promising approach to constrain the opening angle of the dusty torus and to test the unification model. 
Our group investigated 2727 galaxies from the Mapping Nearby Galaxies at Apache Point Observatory (MaNGA) 
and developed spatially resolved techniques for identifying signatures of AGNs \citep{Wylezalek2018}.
A sample containing 308 type II AGN candidates was identified in \cite{Wylezalek2018} via the presence of photo-ionized gas. 
In this paper, we use their narrow-line region morphology to test the AGN unification model. 
One of the challenges is the diverse morphology of these regions, 
which makes it difficult to automate the classification of a cone-like or bi-polar
structures and to measure their orientations and opening angles. 
In this work, we develop a measurement scheme to robustly identify ionization cones and to quantify their morphology, allowing statistical studies of this large sample. 

In the unification model, a correlation between the narrow-line region morphology and the mid-IR color of dust emission is expected because both depend on the inclination of the system. 
When the torus is edge on, the observer can see the cone-like morphology of the narrow-line region with the smallest projected opening angle. 
At the same time, the mid-IR color of this object should be red, because the hotter dust closer to the center is hidden behind the colder one on the outskirts, as predicted by a variety of torus models \citep{Pier1992,Stalevski2012}. 
As the torus becomes more face-on, the observed cone opening angle becomes larger due to projection effects and, eventually, the cones become indistinguishable producing halo-like shape \citep{Mulchaey1996a}. 
The mid-IR colors of these face-on objects are expected to be bluer because of the direct view to the inner hot dust. 
\citet{Fischer2013,Fischer2014} have found correlation between mid-IR color and narrow-line region inclination based on kinematics modeling of bipolar outflows. In this work, we approach this question with resolved two-dimensional NLR morphology. 

Furthermore, with resolved morphology, we can test the relation between the narrow-line region and the host galaxy. The pc-scale dusty torus is not the only obscuring material to block the ionizing radiation of AGN. Structured dust, such as dust lanes, also exists on galactic scales, which have the same effect of obscuring radiation from the nucleus \citep{Lagos2011,Goulding2012} and thus shaping the narrow-line region. 
Therefore, it is important to test the degree of alignment between the narrow-line region and the dusty torus, as well as the alignment between the narrow-line region and the galactic dust. 

The paper is organized as follows. In Section \ref{sec:data}, we describe MaNGA data and supporting multi-wavelength observations. In Section \ref{sec:Measurement}, we describe our morphological measurements. In Section \ref{sec:result}, we present our results on the intrinsic NLR morphology as well as the relationship between the NLR, dust emission, and the host galaxies. We discuss our results in Section \ref{sec:discussion} and conclude in Section \ref{sec:conclusion}. 
Wavelengths in vaccum are used for measurements of the SDSS datasets, but we follow the long-established convention and denote emission lines with their air wavelengths (e.g., [OIII]$\lambda$5007~\AA). Cosmology with $h=0.72,\Omega_{\rm m}=0.3,\Omega_{\Lambda}=0.7$ is adopted throughout this paper. 
We define the inclination of the AGN system (narrow-line region or dusty torus) as the angle between the polar axis and the line-of-sight (LOS). 

\section{Data} \label{sec:data}

\subsection{Sample of MaNGA AGN and AGN narrow line region maps} \label{sec:data:manga}

The Sloan Digital Sky Survey IV (SDSS-IV; \citealt{Blanton2017}) Mapping Nearby Galaxies at Apache Point Observatory (MaNGA;  \citealt{Bundy2014, Drory2014, Law2015, Yan2015, Yan2016}) is an optical fibre-bundle IFU spectroscopic survey conducted with the 2.5 m Sloan Foundation Telescope \citep{Gunn2006} and is one of the three major parts of the ongoing SDSS-IV. 
MaNGA makes use of the Baryon Oscillation Spectroscopic Survey (BOSS) spectrograph \citep{Smee2012}
with a spectral coverage of 3622 -- 10354~\AA~at $R \sim$ 2000.
The diameters of the bundles range from 12 to 32 arcsec, corresponding to 23 to 61 kpc at redshift of 0.1.

Using the outputs from the MaNGA Data Analysis Pipeline, \cite{Wylezalek2018} utilize the traditional [NII] line-diagnostic diagram
\citep{Baldwin1981,Veilleux1987} (hereafter BPT diagram) and construct resolved BPT diagram maps for all galaxies in MaNGA sample.
The [NII]-BPT diagram allows us to distinguish between star-formation, AGN, or Composite (mix of AGN and star formation) dominated emission line regions. From 2727 galaxies observed by MaNGA, \cite{Wylezalek2018} identified a sample of 308 type II AGN candidates, 
which has a redshift range of 0.008 - 0.14 and an \oiii~luminosity (\loiii) range of $10^{39.3} - 10^{41}$ \erg.
In this paper, we focus on the resolved the BPT-maps of this type II AGN sample. 
The spatial resolution of the map is typically 1 - 2 arcsec or 1 - 2 kpc.

\subsection{WISE mid-IR Colors} \label{sec:data:wise}

In order to characterize the effective temperature of the dusty torus, we match our MaNGA AGN sample with the Wide-field Infrared Survey (WISE) photometry data from AllWISE Source Catalog\footnote{\url{http://irsa.ipac.caltech.edu/cgi-bin/Gator/nph-dd}} (W1, 3.4 $\mu m$; W2, 4.6 $\mu m$; W3, 12 $\mu m$; W4, 22 $\mu m$). All sources have WISE matches within 10 arcsec, which is acceptable given the WISE spatial resolution (6--12 arcsec).

In addition to the thermal emission of dust at a wide range of temperatures, the mid-IR spectra of AGN are rich in features, such as silicates at 9.7 and 18 $\mu$m in emission or absorption and polycyclic aromatic hydrocarbon (PAH) features at 6.2, 7.7, 8.6, 11.3, and 12.7 $\mu$m. 
To minimize the impact of these silicate and PAH features, we choose the W2--W4 color as a proxy to the temperature (and thus the inclination) of the dusty torus. 
The wavelength range of the WISE W2, W4 filters are $\sim$ 4--5.5~$\mu$m and $\sim$ 19--28 $\mu$m respectively \citep{Wright2010},
so the W2 and W4 filters avoid most of the spectral features.

The stellar emission in the host galaxy may contribute to the infrared radiation, so we estimate and subtract the stellar component by extrapolating the stellar emission in the optical. 
We obtain the SDSS $i$ (centred at 7625~\AA) and $z$ (centred at 9134~\AA) band photometry (PSF magnitude and Petrosian magnitude) 
for 290 of 308 objects by matching to the position of each object within a circle of 10 arcsec radius and $z < 0.15$
using the SDSS CasJobs\footnote{\url{http://skyserver.sdss.org/casjobs/}}.
We use an Spiral 0 template from the SWIRE template library \citep{Polletta2007} to extrapolate the stellar light to the mid-IR. 
We convert the Spiral 0 template flux into the corresponding magnitude of SDSS $i, z$ and WISE W2, W4 bands.
First, we convolve the template flux with the SDSS and WISE filter transmission 
function of SDSS $\mathrm i, z$ \citep{Doi2010} and WISE W2, W4 \citep{Wright2010} bands to calculate the corresponding flux density.
For SDSS, we use the $\arcsinh$ conversion from \citet{Lupton1999}
to convert the flux into the magnitudes. 
For WISE, we convert the flux into the magnitude using the WISE Zero Magnitude Attributes \citep{Jarrett2011}.
The effective apertures used for deriving WISE magnitudes are likely larger than those of the SDSS PSF magnitudes and smaller than those of 
the Petrosian magnitudes.
So, the PSF magnitude and Petrosian magnitude represent the full range of mid-IR colors corrections (PSF magnitudes underestimate the stellar contribution, Petrosian magnitudes overestimate the stellar contribution).



\subsection{Host Galaxy Morphology from SDSS Catalogs} \label{sec:data:sdss}

To compare the narrow-line region and host galaxy properties, we use SDSS photometry measurements to infer the host galaxy type and orientation. 
We adopt the position angle (\texttt{nsa\_sersic\_phi}), ratio of semi-minor to semi-major axis (\texttt{nsa\_sersic\_ba}) of galaxies from the 2D Sersic fit of the NASA--Sloan Atlas catalog \citep{Blanton2011}. 
In order to distinguish between the disks and the ellipticals galaxies in our sample, 
we use the the fraction attributed to the r-band de Vaucouleurs component (\texttt{fracDeV\_r})
from the SDSS PhotoObjAll table in DR14 and the galaxy zoo fraction of votes for the disk galaxy (\texttt{P\_CS}) from
the Galaxy Zoo 1 data release, 
following the formalism in \citet{Zheng2014} and \citet{Barrera-Ballesteros2016}.
When the \texttt{fracDeV\_r} is less than 0.7 or the \texttt{P\_CS} is greater
than 50\%, this object is considered to be a disk galaxy.
Among our sample of 308 AGN candidates, 225 objects are classified as disks and 60 objects are classified as ellipticals
by this criterion.

\section{Morphological Measurements}
\label{sec:Measurement}

In this section, we describe the narrow-line region map and our methods to identify bi-conical morphology and to quantify its position angle (Sec. \ref{subsec:bipolar}), opening angle (Sec.~\ref{sec:opening}), and concentration (Sec. \ref{sec:concentration}).

\subsection{Identification of Bi-polar Ionization Cone}
\label{subsec:bipolar}

\begin{figure*}
\center{}
\includegraphics[height=7.cm,angle=0]{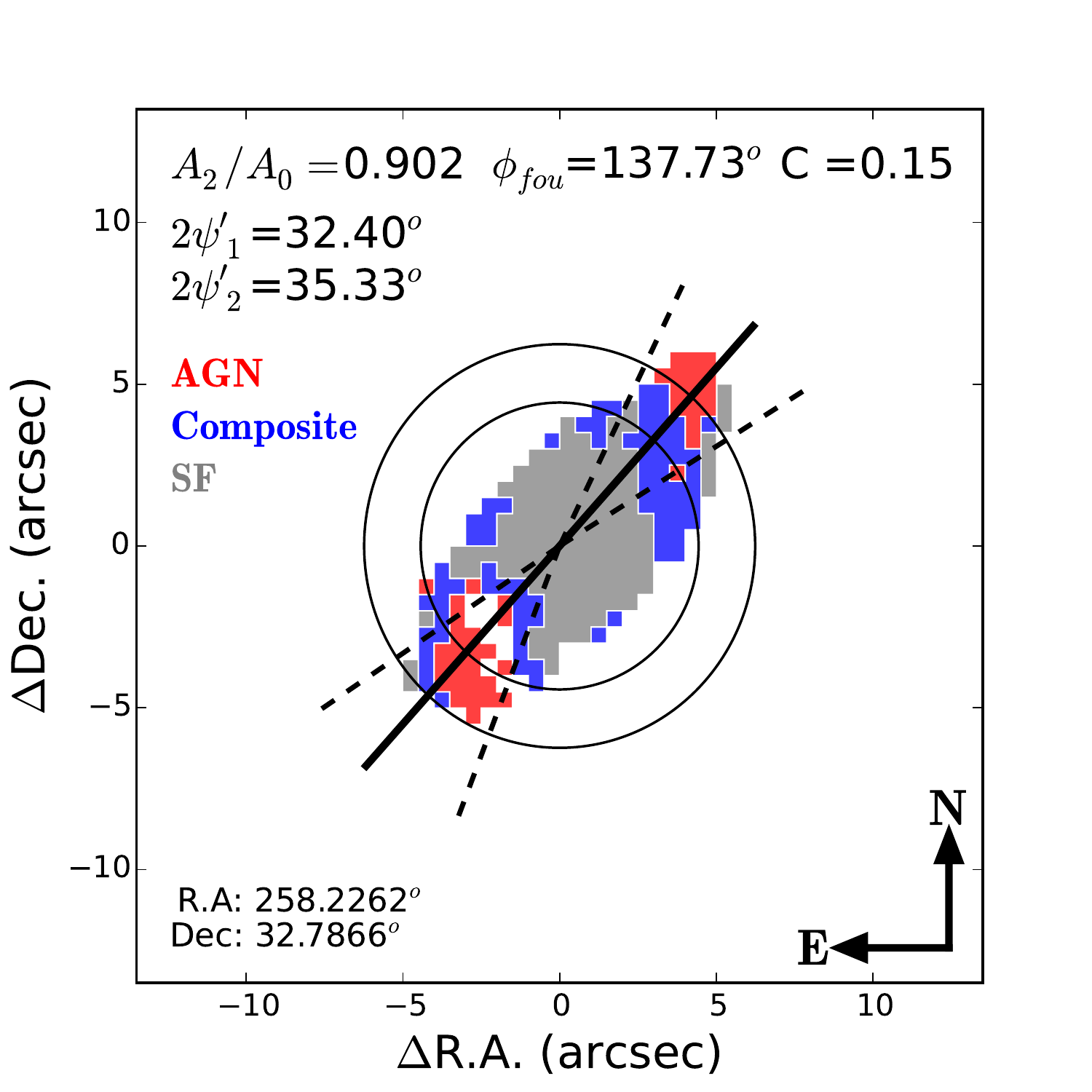}
\includegraphics[height=7.cm,angle=0]{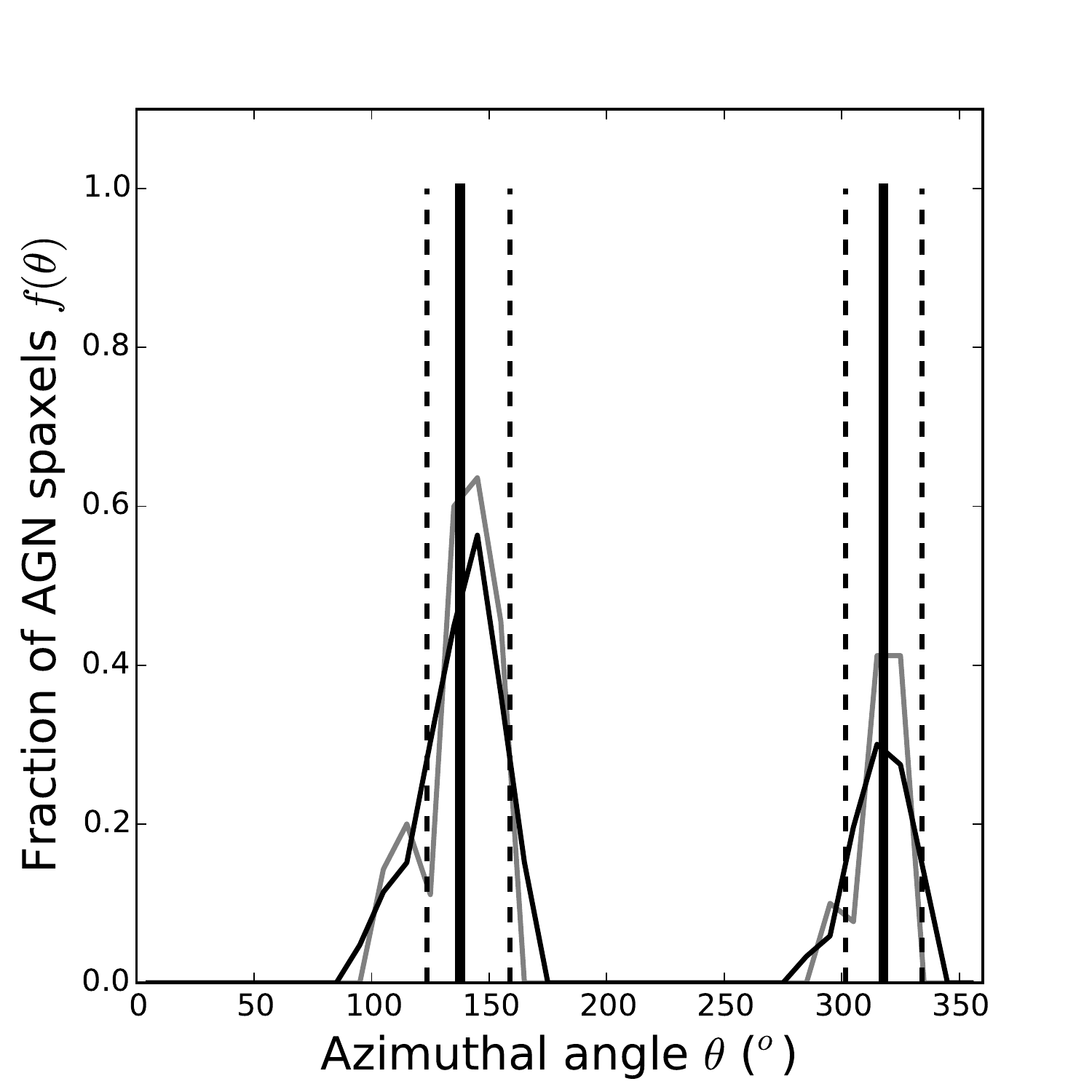}
\includegraphics[height=7.cm,angle=0]{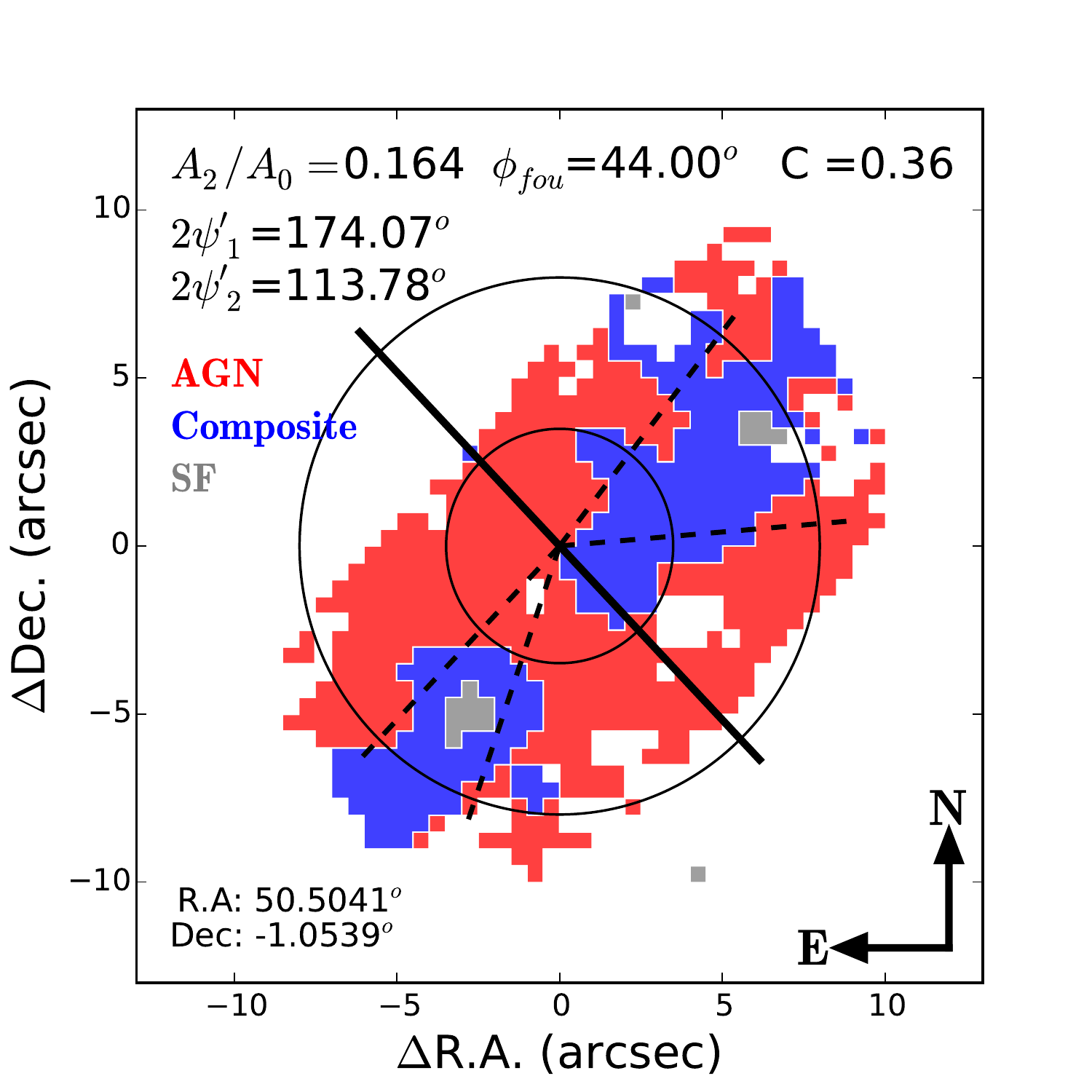}
\includegraphics[height=7.cm,angle=0]{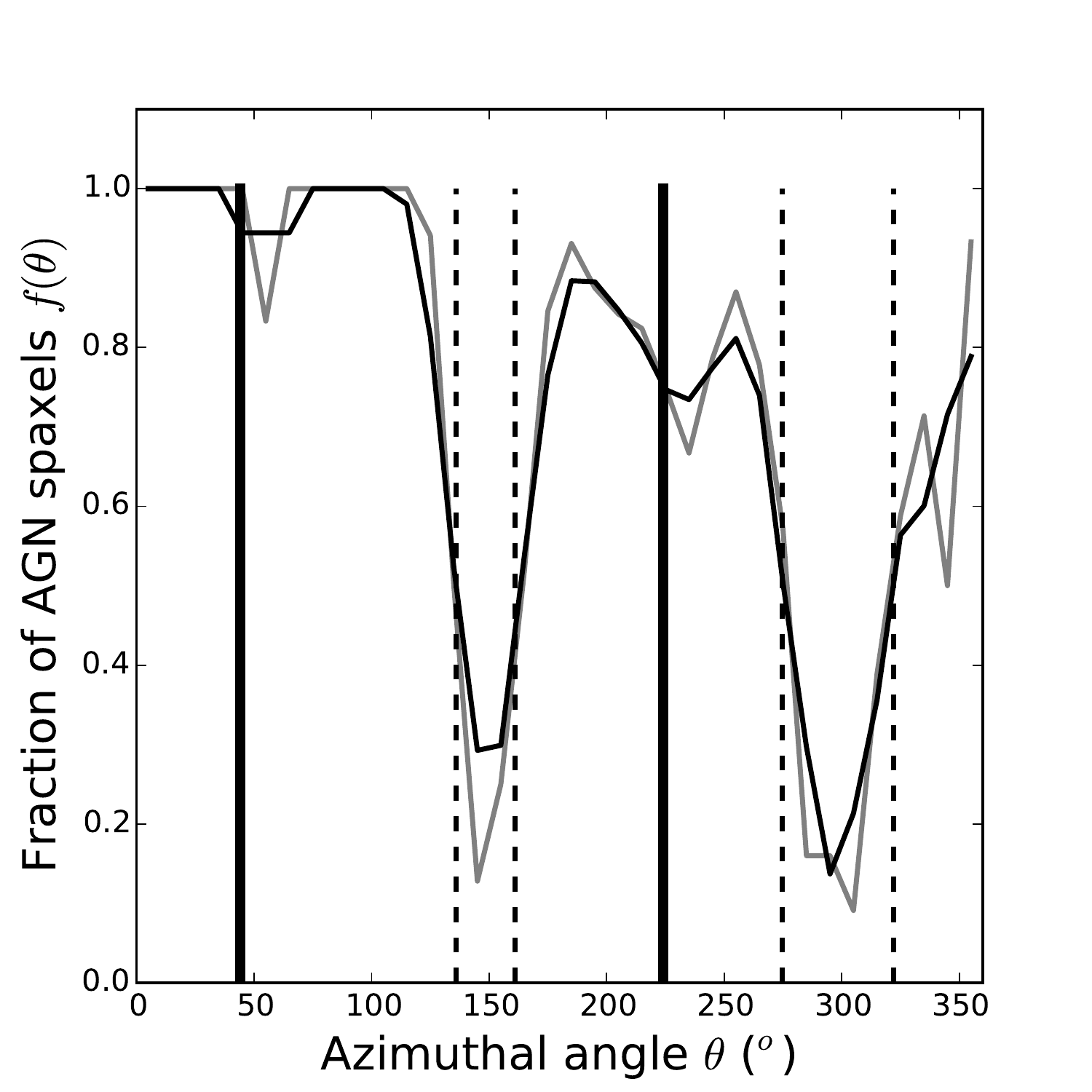}
\includegraphics[height=7.cm,angle=0]{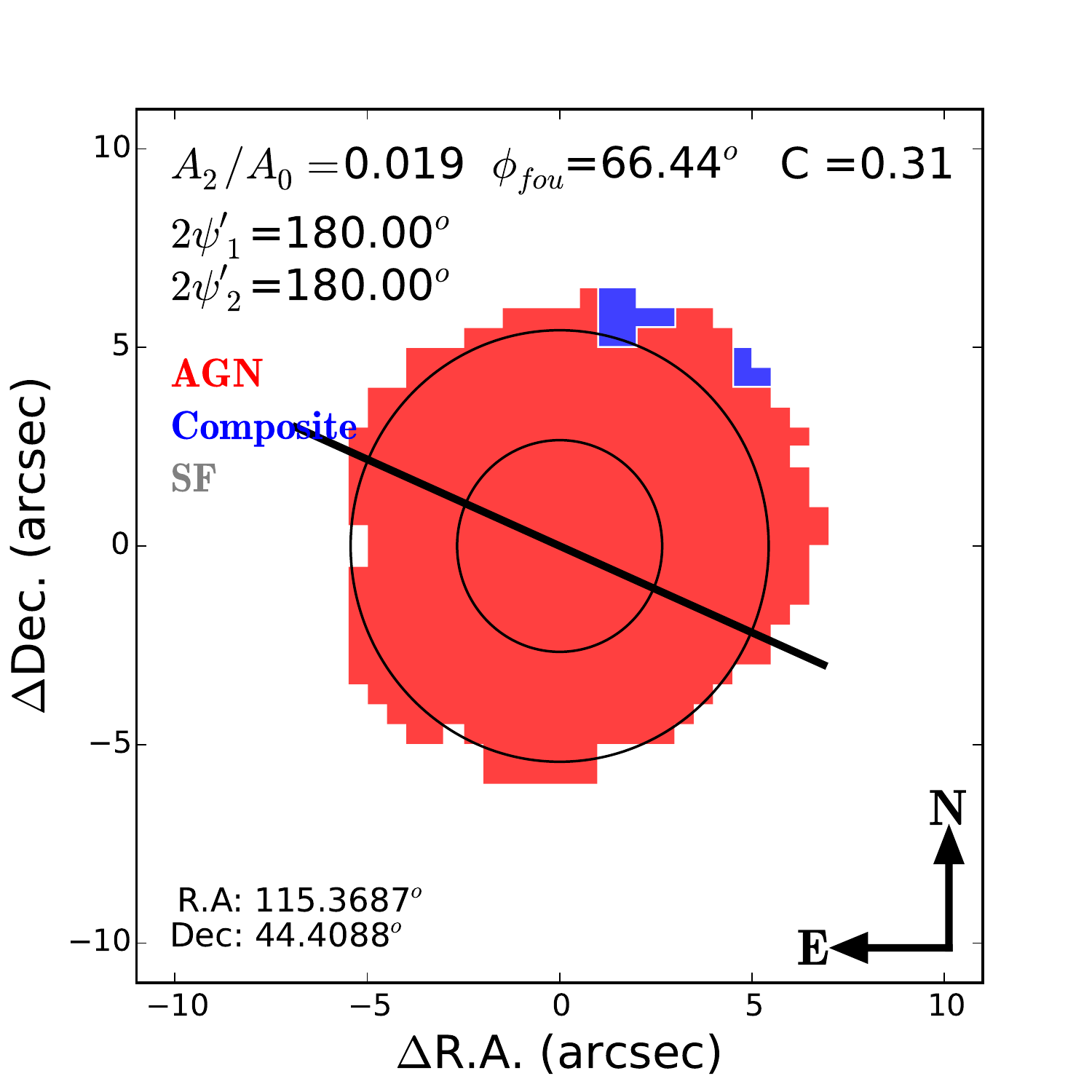}
\includegraphics[height=7.cm,angle=0]{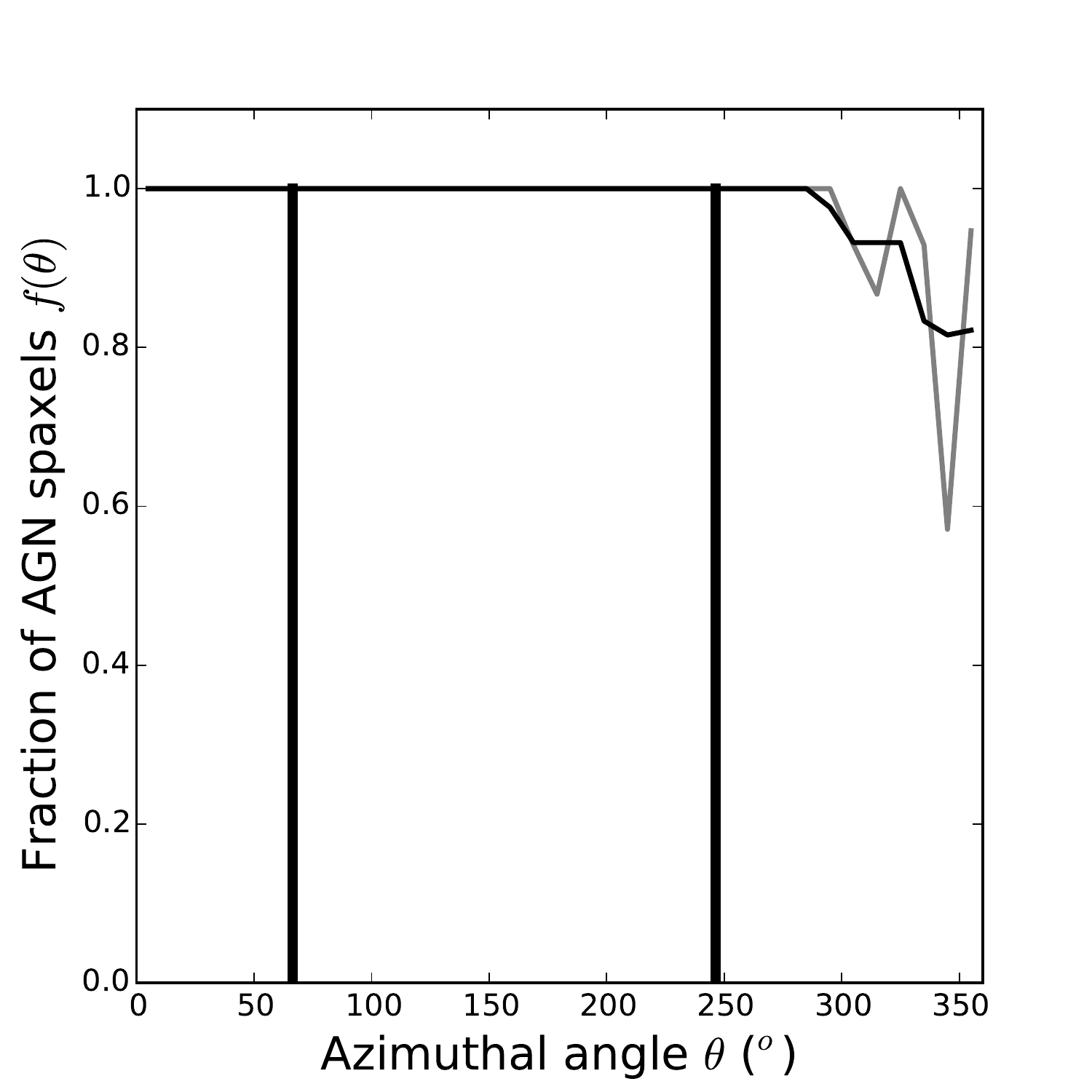}
\caption{Examples of measuring the position angle, opening angle and concentration. Left panels are the [NII]-BPT diagrams. The gray curves in right panels are the corresponding fraction of AGN spaxels along the azimuthal angle. The black curves are produced by smoothing the gray ones with a three-pixel boxcar. The position angle (black lines) is the result of our Fourier analysis. Boundaries (dashed lines) of bicone are determined by where the AGN fraction drops to half of the maximum value. $A_2/A_0$ is the ratio of mode $m=2$ to $m=0$ of the Fourier series. $C$ is the concentration index of the emission-line region: the logarithmic ratio of the circular radius containing 80\% ($r_{80}$, black circle) of a cone's light to the radius containing 20\% ($r_{20}$, black circle) of the AGN spaxels.}
\label{fig2}
\end{figure*}

In Fig.~\ref{fig2}, we show several example AGN candidates from \citet{Wylezalek2018}. As shown in the left panel, each spaxel is classified as either star-fromation-dominated, composite, or AGN-dominated based on its position in the BPT diagram \citep{Wylezalek2018}. 
For the purpose of measuring the narrow-line region morphology, we take only the AGN spaxels as part of the narrow-line region. These regions display a diverse range of morphology -- disconnected blobs, bicone, centralized blob, ring, irregular, or some combination of the above. Our final measurements are based on an ionization bit map, in which spaxels are set to 1 if they are classified as AGN-dominated and 0 otherwise.
As shown in the maps, the data is confined to a region where there is high signal-to-noise ratio (S/N). 
This creates an arbitrary mask on the data which has to be taken into account in the morphological measurements. 

The bi-conical morphology is a strong periodic feature when expressed in polar coordinates. 
Specifically, because it appears twice in the 2$\pi$ circle, it corresponds to the $m=2$ mode in the azimuthal Fourier series representation of the ionization bit map. 
To measure the strength of this mode, we express the ionization bit map as a function of the azimuthal angle, $\theta$, as shown in the right panel of Fig.~\ref{fig2}, and to decompose it into a Fourier series.  

The signal-to-noise mask imposes a cut-off to the number of spaxels in each azimuthal angle. So instead of counting the total number of AGN spaxels along each direction, we use the fraction of AGN spaxels, $f(\theta)$. 
The circle is divided into 36 bins each 10 degrees wide. The AGN spaxel fraction $f(\theta)$ is calculated in each of the bins and expressed as a function of the East of North position angle $\theta$. 
The origin is taken as the center of the IFU, which coincides with the photometric center of the galaxy measured from SDSS image.

We then decompose the fraction as a function of position angle $f(\theta)$ into a Fourier series:
\begin{eqnarray} \label{eq01}
f(\theta_{k}) = a_0 + \sum_{m=1}^{\infty} [a_m \cos(m\theta_{k}) +  b_m \sin(m\theta_{k})],
\end{eqnarray}
where
\begin{eqnarray}\label{eq02}
\left\{  
\begin{array}{rcl}  
a_m & = & \frac{2}{2\pi}\sum_{k=1}^{36}  f(\theta_{k})\cos(m\theta_{k})
\frac{10\pi}{180}  \\
&  &  \\
& = & \frac{1}{18}\sum_{k=1}^{36} f(\theta_{k})\cos(m\theta_{k}), \\
&  &  \\
b_m & = & \frac{1}{18}\sum_{k=1}^{36} f(\theta_{k})\sin(m\theta_{k}), \\
&  &  \\
\theta_{k} & = & (10k-5)~deg; (k = 1, 2, 3, ..., 36). \\

\end{array}
\right.
\end{eqnarray}

The $m=0$ mode represents the circularly symmetric component and $m=2$ the bi-polar component. We use the ratio of the amplitudes of the $m=2$ to the $m=0$ modes, $A_2/A_0 = \sqrt{a_2^2+b_2^2}/a_0$, to represent the significance of the bi-polar component. 
The $A_2/A_0$ value is in the range of 0 to 1. A small value means that the morphology is close to circularly symmetric (e.g., bottom right of Fig.~\ref{fig2}) while a large value represents a more bipolar or biconical morphology (e.g., top right of Fig.~\ref{fig2}). 
We visually inspect each object and find that most objects with $A_2/A_0$ above 0.1 show clear bicone morphology, which accounts for 248 (80.5\%) objects among the sample. This high fraction supports that majority of the narrow-line regions should be intrinsically bi-polar.

When the $m=2$ mode is present, we can use the phase of this Fourier mode as the position angle of the 
bicone, $\phi_{fou} = \arctan (b_2/a_2)/2$, shown in Fig.~\ref{fig2}.
The objects with lower values of $A_2/A_0 < 0.1$ do not exhibit bicone morphology and their position angle $\phi_{fou}$ becomes ill-defined. 
Hereafter, we use the ratio $A_2/A_0$ as a measure of the strength of the biconical morphology and $\phi_{fou}$ as its position angle.

\subsection{Opening Angle of Ionization Cone}
\label{sec:opening}

The opening angle is an important characteristic of the bicone, which, in the unification model, is tied to the ratio of type I to type II AGN. However, with the 2D NLRs map, the only observable is the projected opening angle 2$\psi'$ ($\psi'$ is the projected half opening angle)
on the plane of the sky. 

When a bicone is present, the edges of the cone correspond to sudden drops in the fraction of the AGN spaxels along a certain azimuthal angle. 
To capture this feature, we define the cone boundary to be where the AGN fraction curve, $f(\theta)$, drops to half of its maximum value, i.e., the opening angle is the full width at half maximum (FWHM) of the fraction curve. 
To mitigate noise and the non-monotonic features in the curve, we smooth the angle function $f(\theta)$ with a three-pixel boxcar. 
If the boundaries cannot be found, the opening angle is set to 0 deg when the mean value is lower than half of the maximum value, otherwise, it is set to 180 deg. 
The opening angle of the bicone on each side is measured separately. Their mean is taken to be the representative projected opening angle for the AGN. 
The measured position angles and opening angles are in excellent agreement with the by-eye classification.

As shown in Fig.~\ref{fig3}, there is a strong anti-correlation (Spearman correlation 
Test: $r = -0.55$,~$p$-value $ < 10^{-9}$, listed in Table.~\ref{table1}) between the strength of the bicone $A_2/A_0$ and its opening angle. 
This is consistent with our expectation that when the projected opening angle becomes large, the bicone becomes less distinct and prominent.

\subsection{Concentration}
\label{sec:concentration}

The inclination of the bicone also affects the radial distribution of the narrow-line region map. 
When viewed pole-on, the bicones would have a centrally concentrated morphology. When the inclination angle is larger, the V-shaped morphology is less radially concentrated \citep{Mulchaey1996a}.

The concentration index $C$ \citep{Bershady2000} was first adopted for stellar light to describe the morphology of galaxies
-- high concentration values are characteristic of ellipticals and lower ones of disks.
In this paper, we use it to describe the distribution of the AGN emission region. 
It is defined as the logarithmic ratio of the circular radius containing 80\% ($r_{80}$) of the AGN spaxels to the radius containing 20\% ($r_{20}$) of the spaxels, i.e. $C = log_{10} (r_{80}/r_{20})$. 
A large concentration value indicates a majority of AGN region is concentrated at the center.
For a solid circle, $C = log_{10} (2) \approx 0.3$. If the center is not filled, e.g. a ring (Fig.~\ref{fig4}), or a bicone, the value of $C$ is smaller than 0.3. 
Through the visual inspection, we find that the morphology of AGN regions in 9 objects that have small values of $C$ is ring-like.

\begin{figure}
\center{}
\includegraphics[height=7.1cm,angle=0]{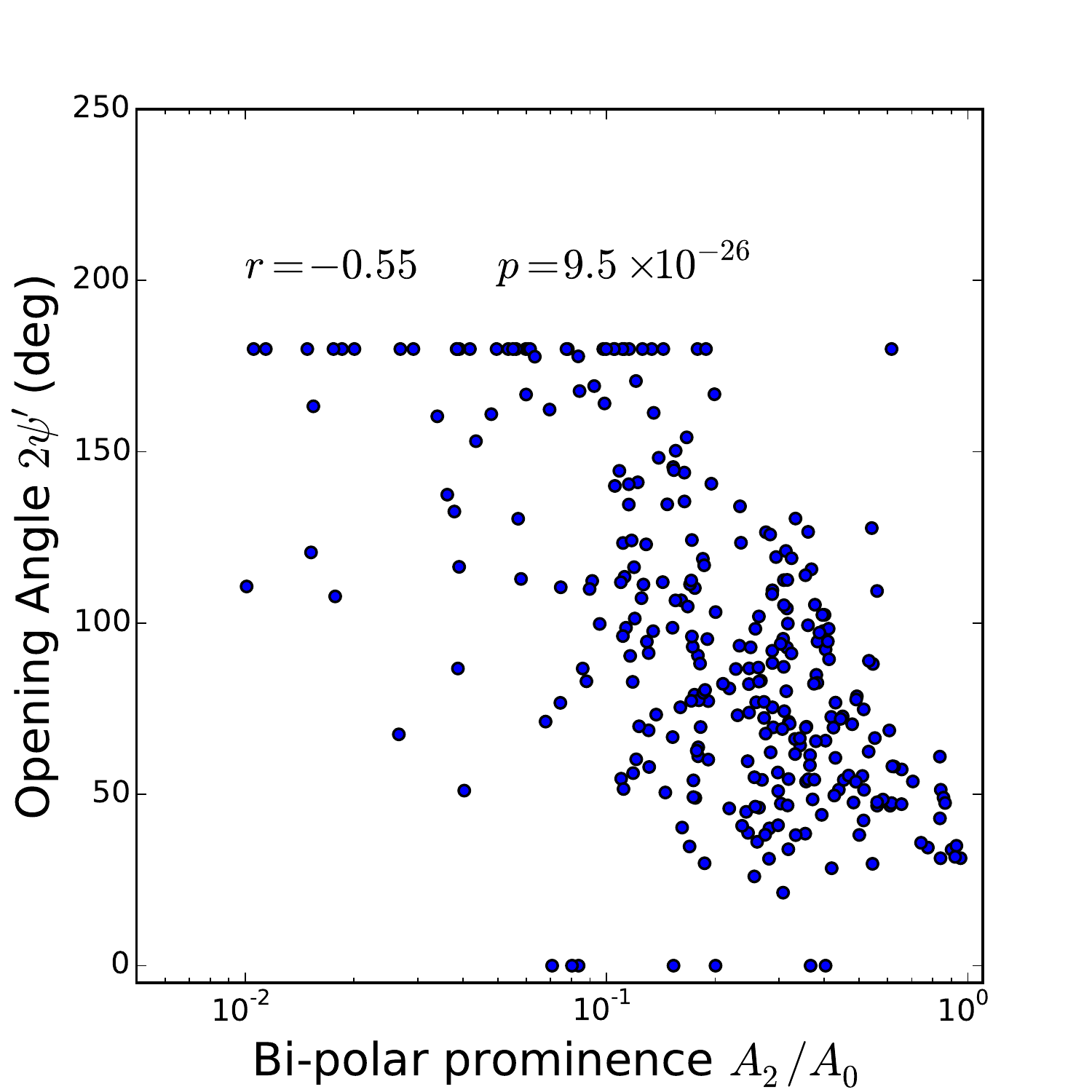}
\includegraphics[height=7.1cm,angle=0]{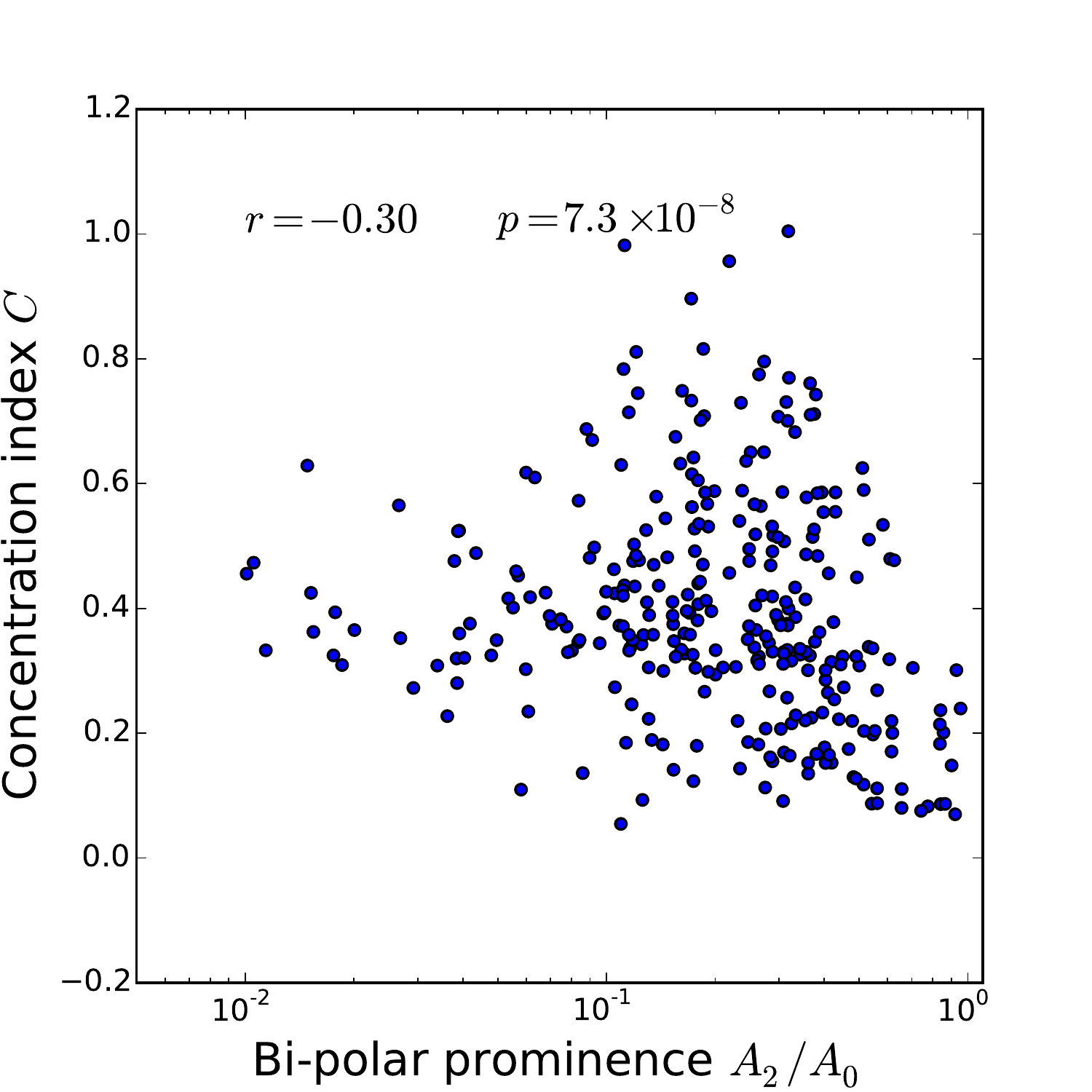}
\includegraphics[height=7.1cm,angle=0]{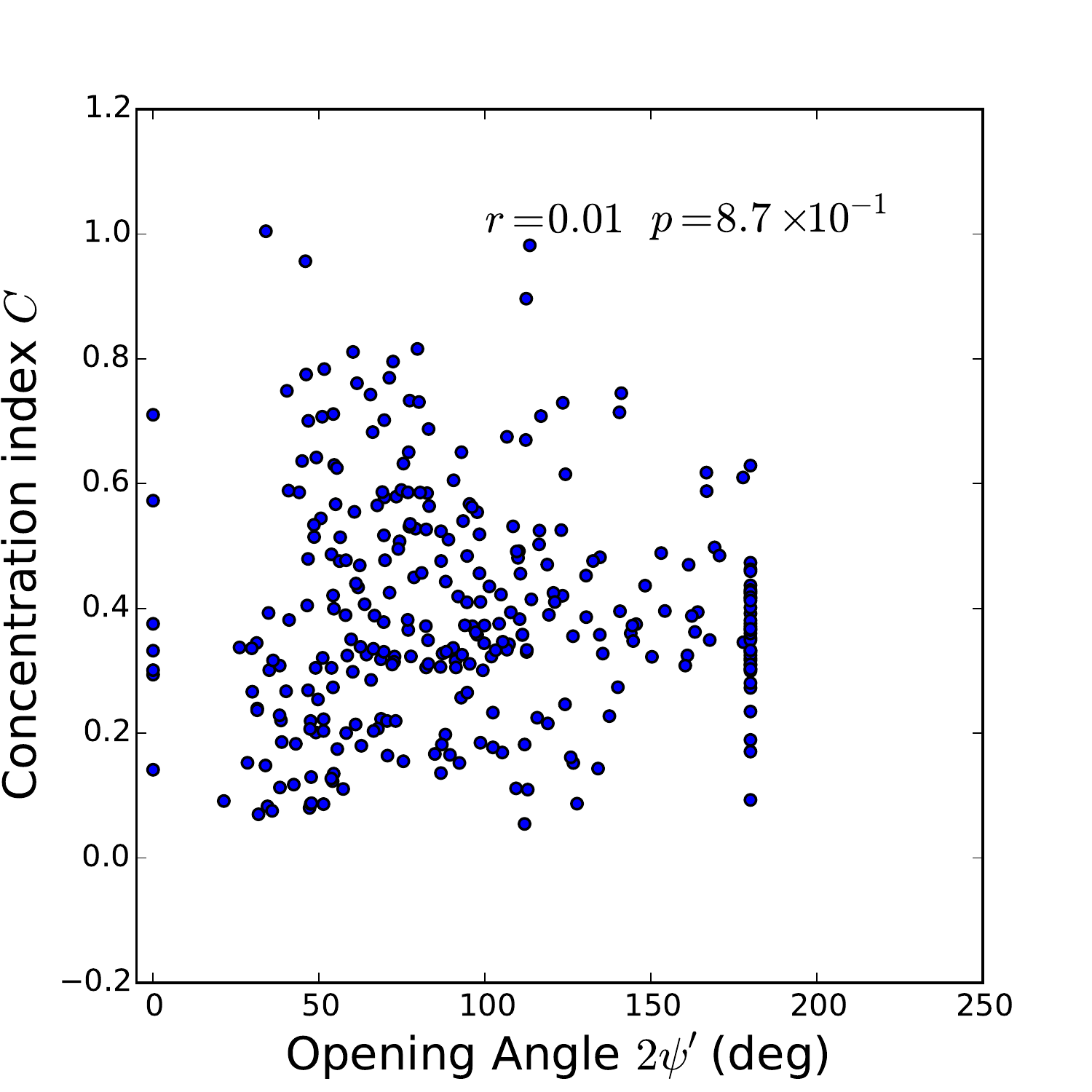}
\caption{The relationships between the $A_2/A_0$, opening angle and concentration index $C$. The rank correlation coefficient $r$ and the probability of the null hypothesis (no correlation) $p$ are indicated in every panel.}
\label{fig3}
\end{figure}

\begin{figure}
\center{}
\includegraphics[height=6.7cm,angle=0]{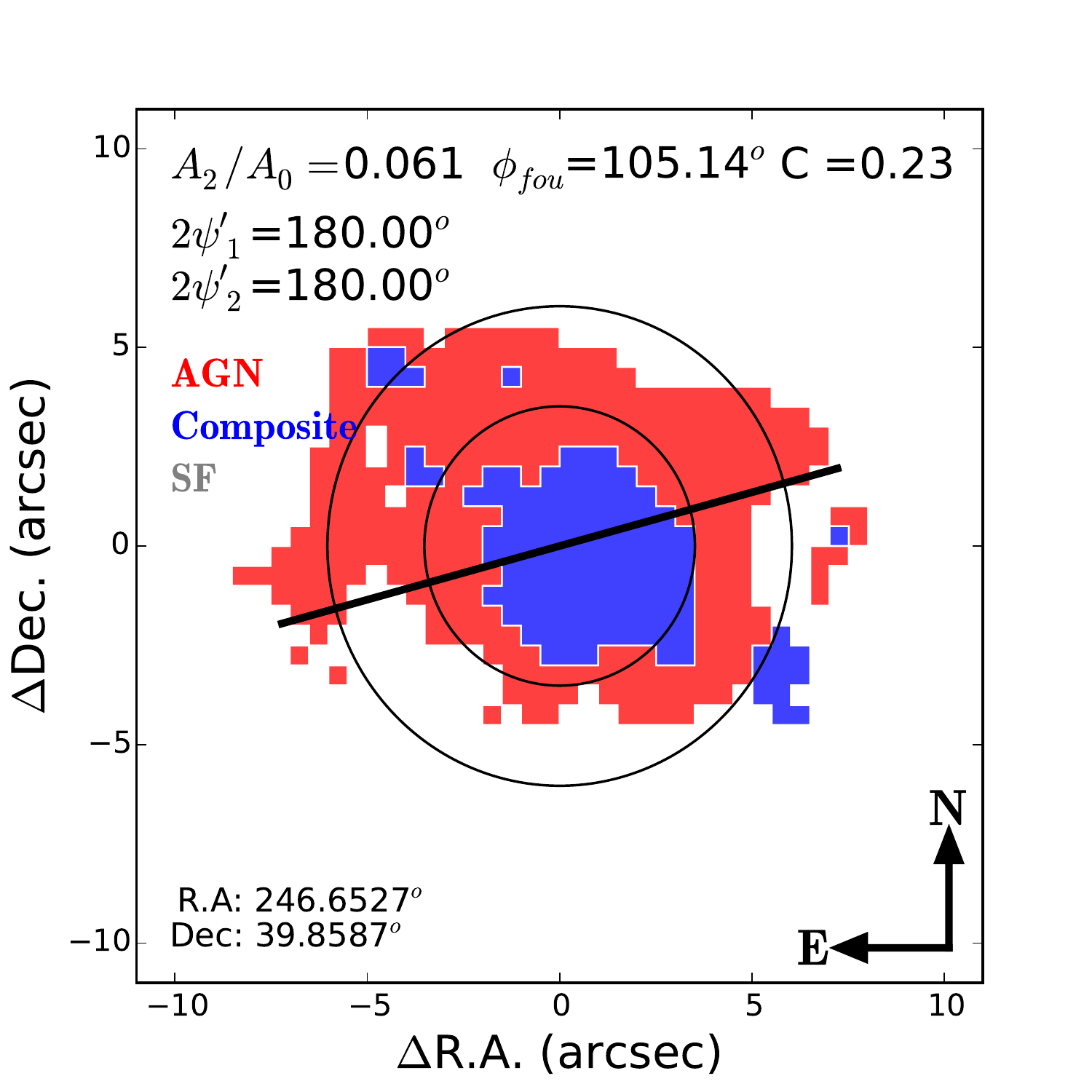}
\includegraphics[height=6.7cm,angle=0]{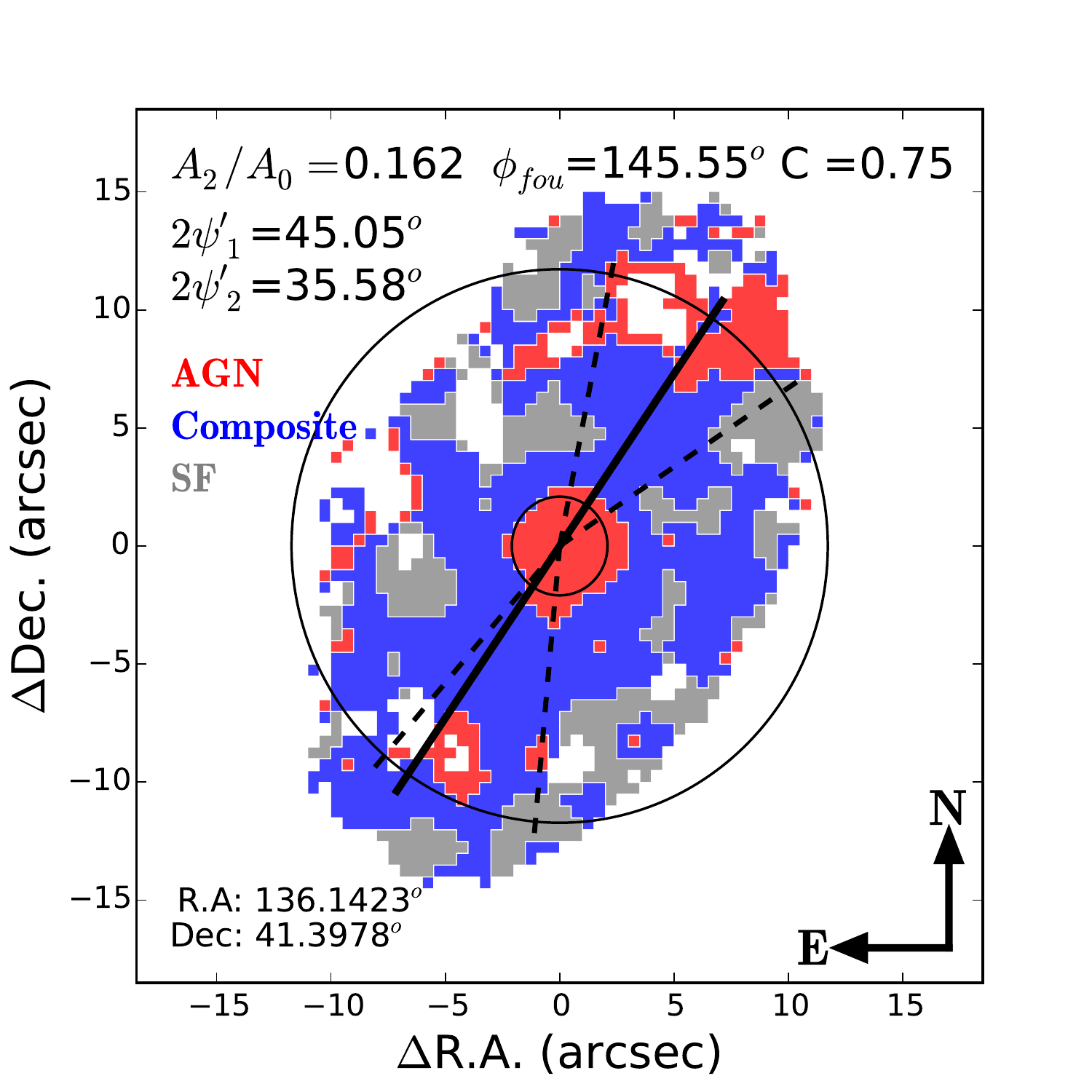}
\caption{Examples of AGN with small and large concentration values.}
\label{fig4}
\end{figure}

As shown in Fig.~\ref{fig3}, there is a significant anti-correlation ($r = -0.3$,~$p$-value $=  7.3 \times 10^{-8}$) between the $A_2/A_0$ and $C$ index. 
The correlation between the opening angle and $C$ is not significant ($r = 0.01$,~$p$-value $=  0.87$). 
This indicates that systems with more prominent bicones (larger values of $A_2/A_0$) are less concentrated, possibly because of the high inclination angle of the cones, consistent with our expectations. 


\section{Results}
\label{sec:result}

In this section, we present our analysis and results on the intrinsic opening angle of the ionization cone (Sec. \ref{sec:intrinsic}), the correlations between the narrow-line region morphology and the mid-IR color (Sec. \ref{sec:infrared}), and the alignment between the narrow-line region and the host galaxy (Sec \ref{sec:result:align}). 

\subsection{The Intrinsic Opening Angle of Ionization Cones}
\label{sec:intrinsic}

\begin{figure*}
\center{}
\includegraphics[height=7.5cm,angle=0]{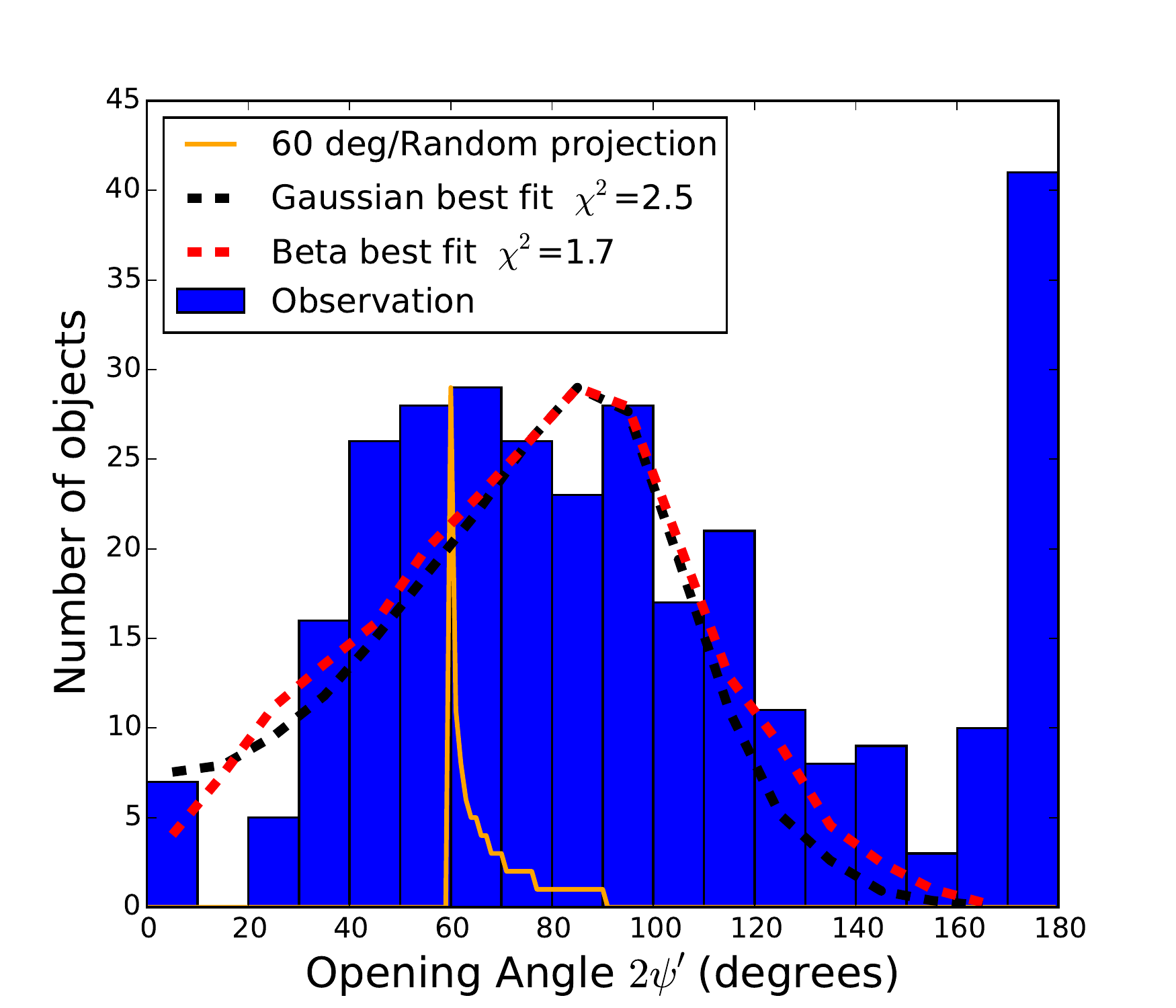}
\includegraphics[height=7.5cm,angle=0]{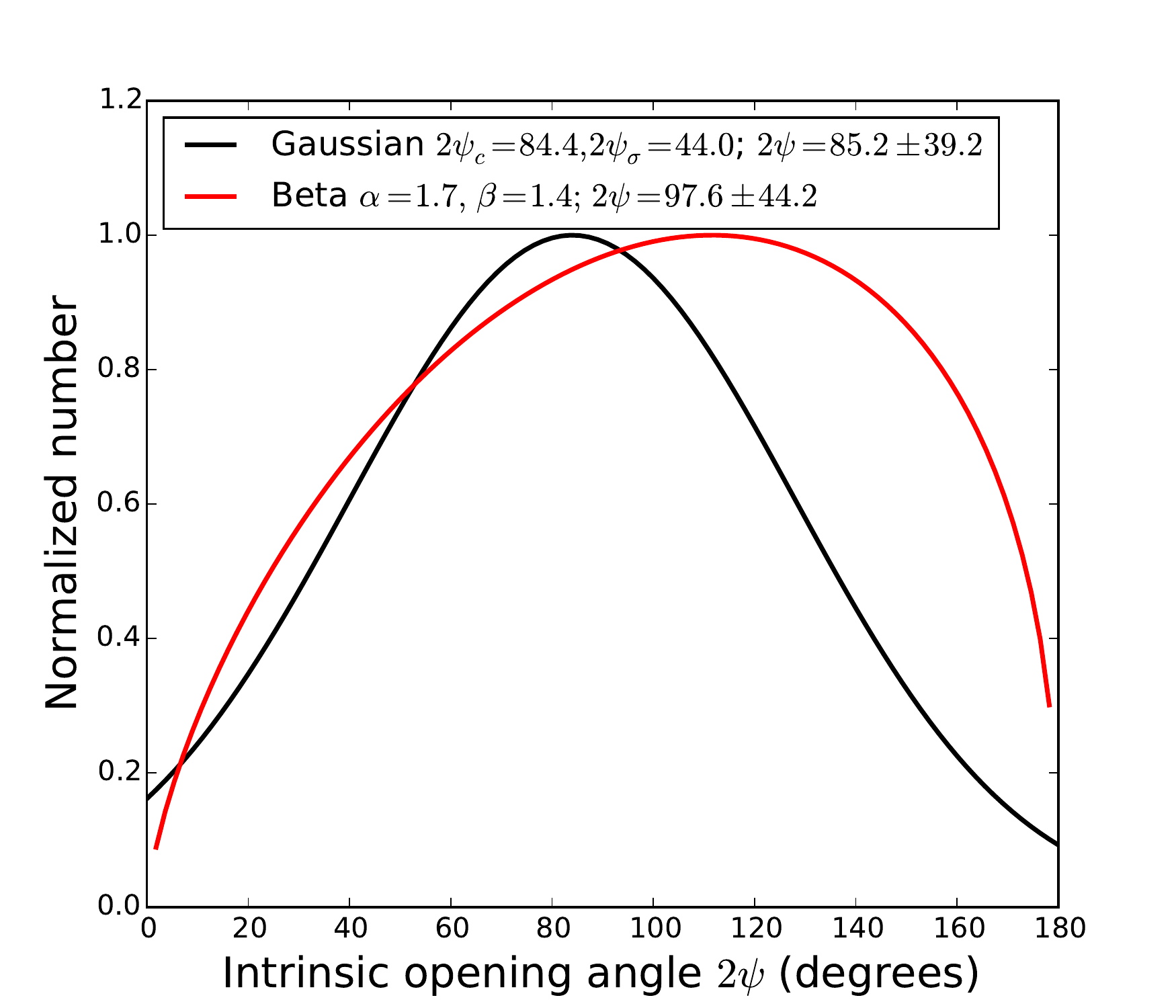}
\includegraphics[height=8.0cm,angle=0]{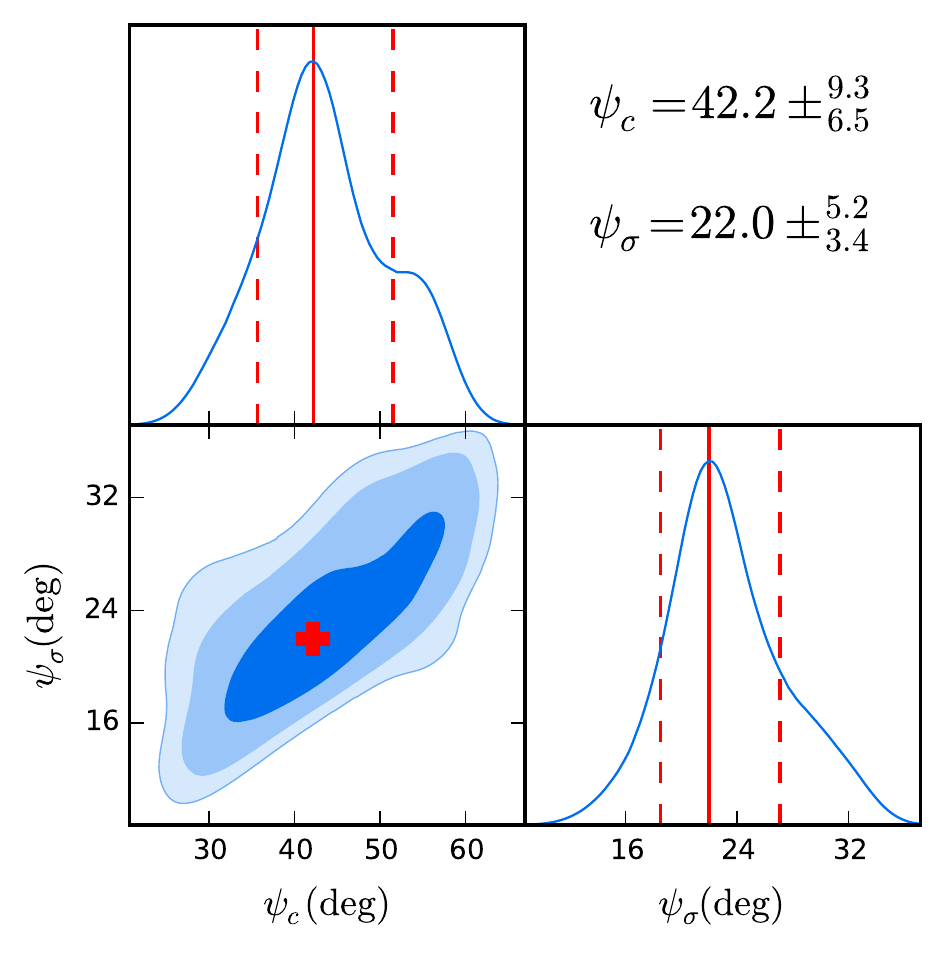}
\includegraphics[height=8.0cm,angle=0]{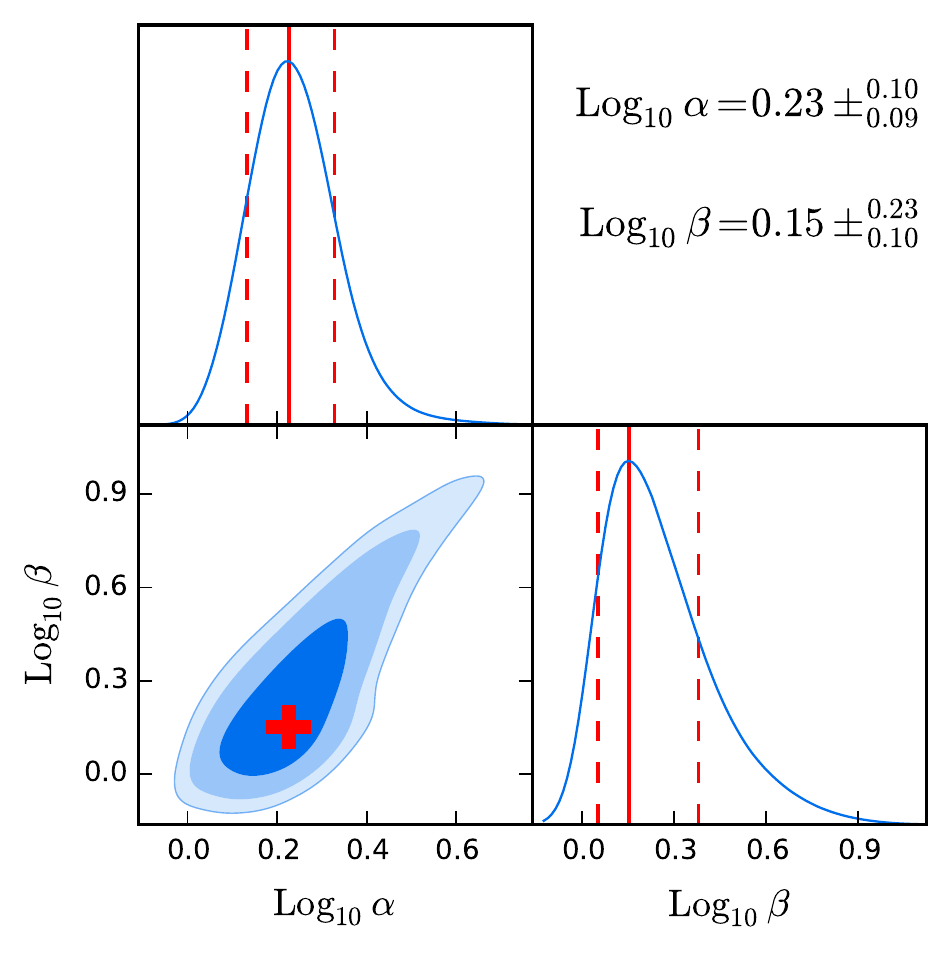}
\caption{The MCMC result of the distribution of opening angles. We assume a Gaussian and a Beta distribution of the intrinsic angle 
projected randomly and isotropically on to the plane of the sky. 
Upper left: The data and best-fit model. The vertical red solid line represents the peak value of the intrins. Upper right: the best fit 
intrinsic angle of Gaussian and Beta function. Lower panel: the posterior on the Gaussian G($\psi_c, \psi_{\sigma}$) and 
Beta B($\alpha, \beta$) parameters of the intrinsic half opening angle distribution. The red cross marks the best-fit value of the parameters. The red solid (dashed) lines represent the peaks ($1 \sigma$ error) of the parameters.}
\label{fig6}
\end{figure*}

A large fraction (80.5\%) of narrow-line regions in type II AGN in our sample show biconical morphology, in agreement with the prediction of the unification model. 
From the distribution of the projected opening angles (Fig.~\ref{fig6}), 
we can infer the distribution of the bicone intrinsic opening angles.
In our analysis, we must take into account the selection effect that the MaNGA AGN sample from \citet{Wylezalek2018} contains only type II AGN. Modulo this constraint, we make an assumption that the viewing angle is random. 

First, we determine the expected distribution of projected angles in the case of a single cone opening angle. 
For a given cone with an intrinsic half opening angle of $\psi$ at an inclination angle of $\theta$, the projected half opening angle $\psi'$ can be expressed as:
\begin{eqnarray} \label{eq03}
\psi' =& \arctan\left( \frac{\tan~\psi}{\sin~\theta}\right)~ \cdots \cdots~ \theta > \psi, \\
\psi' =& 90 ^\circ ~\cdots \cdots~ \theta \leq \psi. 
\end{eqnarray}
Here the inclination angle $\theta$ is the angle between the line-of-sight and the bicone axis. 
If the inclination angle $\theta$ is smaller than $\psi$, i.e., the observer's line-of-sight is inside the cone (type I AGN), then we discard the system from the predicted distribution because it would be excluded from our observed sample based on our type II AGN selection. 

As an illustration, the distribution of isotropically projected opening angles of a cone with a fixed half opening angle $\psi$ is shown as the orange line in up left panel of Fig \ref{fig6}. 
It is narrow and peaked around the value of the intrinsic opening angle, because there is a relatively high chance of being close to edge-on ($\theta \sim$ 90 degrees). 
In fact, this distribution is much narrower than the observed distribution of the projected half opening angles $\psi'$, which makes it clear that there has to be a range of intrinsic opening angles in order to explain the observations. 

We assume that such a distribution of the intrinsic half opening angle is a smooth function that can be approximated by analytical models. To include possibilities of different dispersion and skewness in the distribution, we consider Gaussian and Beta distribution. 
The Gaussian $P(\psi) = G(\psi_c,\psi_\sigma)$ is a symmetric distribution parametrized by the central value $\psi_c$ and the standard deviation $\psi_\sigma$. 
The Beta distribution $P(\psi) = B(\alpha,\beta)$, parametrized by $\alpha$ and $\beta$, is assymetric if $\alpha \neq \beta$. 
The parameters are to be determined by fitting the model predicted distribution to our data with a Markov chain Monte Carlo (McMC) procedure described in the following:

Given an intrinsic half opening angle distribution $P(\psi) = G(\psi_c,\psi_\sigma)$ or $P(\psi) = B(\alpha,\beta)$, we predict the projected opening angle distribution $P(\psi')$ with an Monte Carlo simulation:

\begin{itemize}

\item[1.] We draw 200 intrinsic opening angles $\psi$ from the Gaussian distribution $G(\psi_c,\psi_\sigma)$ or $B(\alpha,\beta)$. 

\item[2.] For each of the intrinsic opening angle $\psi$, we project it, according to Eq. \ref{eq03}, with 500 realizations of random isotropic inclination angles $P(\theta) \propto \cos(\theta)$. 

\item[3.] We compile the distribution of 200 $\times$ 500 = 100,000 projected opening angles, discard the ones corresponding to type I AGN ($\theta < \psi$), and then normalize the distribution. 

\end{itemize}

We then fit this predicted opening angle distribution to the data using a McMC package (\verb emcee ,~\url{http://dfm.io/emcee/current/user/line}). 
The likelihood is evaluated for each data point, i.e., each opening angle bin of $10^\circ$, 
based on an assumed Gaussian likelihood function. 
The total likelihood is a product of all the likelihoods in all the bins. 
We adopt flat priors on the parameters and run $10^5$ steps in the Markov chain.

The bin with the largest projected opening angle 2$\psi' = 180$ is excluded in the fit, which accounts for a small fraction of the sample (12\%, 38/308). 
Systems in this bin corresponds to halo-like emission line regions with no clear bi-cones. 
In unification model, these systems should correspond to type I AGN and would not be selected in our sample. 
But it is possible that beam smearing limits our ability to resolve bi-cones in some cases. 
Alternatively, the dust geometry in these cases may be more complex than assumed in the dusty torus model. 
Either way, these cases account for a small fraction of the sample. In our analysis, they are excluded in the modeling. 

By construction, there is not much constraining power on the population of AGN with the largest opening angles. This is because such systems are most likely observed as type I AGN and would contribute little to the type II AGN population even if they are abundant. 
On the other hand, the population with small opening angles should be relatively well-constrained. 
The number of AGN with small projected opening angle is relatively small in our data. 
So although there are uncertainties on the projected opening angles at the $10-20^\circ$ level due to the finite bin size and smoothing, etc., such that the number count in each bin may not be exact, our data implies that there is no significant population of AGN with small intrinsic cone opening angles. 

The posterior distribution of the Gaussian parameters ($\psi_c$, $\psi_\sigma$) and Beta parameters ($\alpha,\beta$) are shown in Fig.~\ref{fig6}.
For the Gaussian function, the best-fit values (peak of the posterior distribution) and the errors (68\% credible interval) of the peak half opening angle and the standard deviation are $\psi_c = 42.2\pm^{9.3}_{6.8}$ deg and $\psi_\sigma = 22.0\pm^{5.2}_{3.4}$ deg, respectively. 
For the Beta function, the best-fit values (peak of the posterior distribution) and the errors (68\% credible interval) are $\log_{10}\alpha = 0.23\pm^{0.10}_{0.09}$ and $\log_{10}\beta = 0.15\pm^{0.23}_{0.10}$, respectively. 
The peak of the intrinsic opening angle and its standard deviation is 2$\psi = 84.4\pm 44.0$ deg for Gaussian function and 2$\psi = 97.6\pm 44.2$ deg for the Beta funciton. 
These fits reasonably capture the peak and the spread of projected opening angles, but there is some small excess and deficit at angles that is not captured in the model, possibly indicating that the intrinsic distribution $P(\psi)$ differs from the assumed ones. 

The best-fit Gaussian and Beta functions are consistent at small intrinsic opening angles but deviate at larger angles. 
This highlights the lack of constraining power in those regimes in the sense that a dramatic increase in the population with large opening angles would not change the observables singnificantly due to our selection. 
This uncertainty would result in uncertainties in the contrained type ratio. 
The Gaussian and Beta functions give different type II fractions. For the Gaussian distribution, the type II fractions is $\approx~70\%$, corresponding to a number ratio between type I and type II AGN of $\approx~1 : 2.3$. 
For the Beta distribution, the type II fractions is $\approx~61\%$, corresponding to a number ratio between type I and type II AGN of $\approx~1 : 1.6$. 
This is calculated by averaging the type II fraction $\langle \cos \psi \rangle$ with one million realizations of $\psi$ drawn from 
the best-fit Gaussian and Beta distributions $P(\psi)$.
Its implication and comparison with the literature are discussed in Sec. \ref{sec:discussion:ratio}.

\subsection{Correlation between NLR morphology and mid-IR color of dusty torus}
\label{sec:infrared}

As discussed in the introduction, a correlation between the NLR morphology and the mid-IR color of AGN is expected because both of them depend on the inclination angle of the system under the framework of the unification model. 
When the observer views the dusty torus from the edge, the mid-IR emission should be redder (high W2 - W4) than in the face-on case due to the colder effective dust temperature  of the outer parts of the torus, as predicted by various (clumpy or smooth) torus models. At the same time, the narrow-line region should be seen as prominent bicones (high $A_2/A_0$), with smallest opening angle (low 2$\psi'$), and low concentration (low $C$). 

\begin{figure*}
\center{}
\includegraphics[height=7.9cm,angle=0]{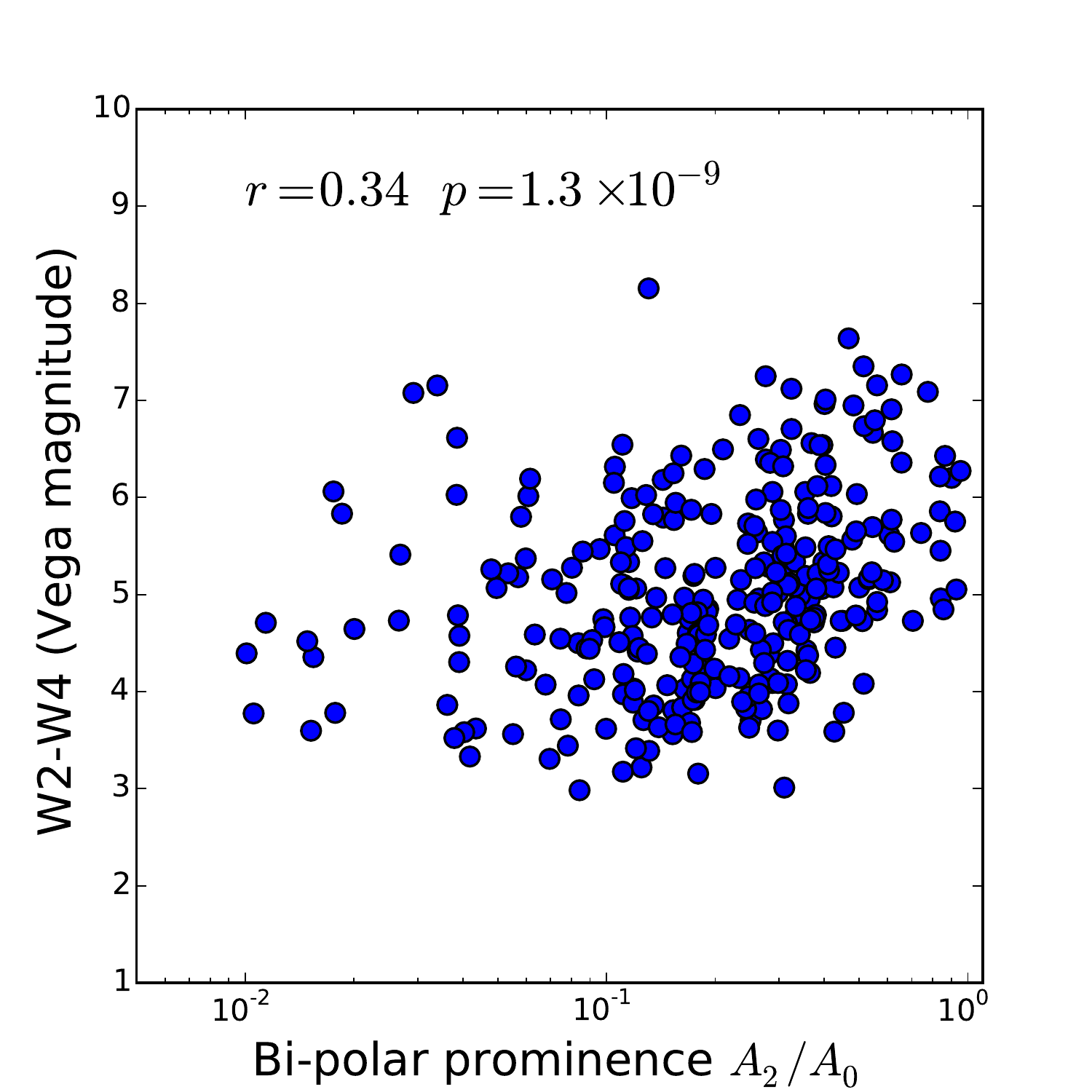}
\includegraphics[height=7.9cm,angle=0]{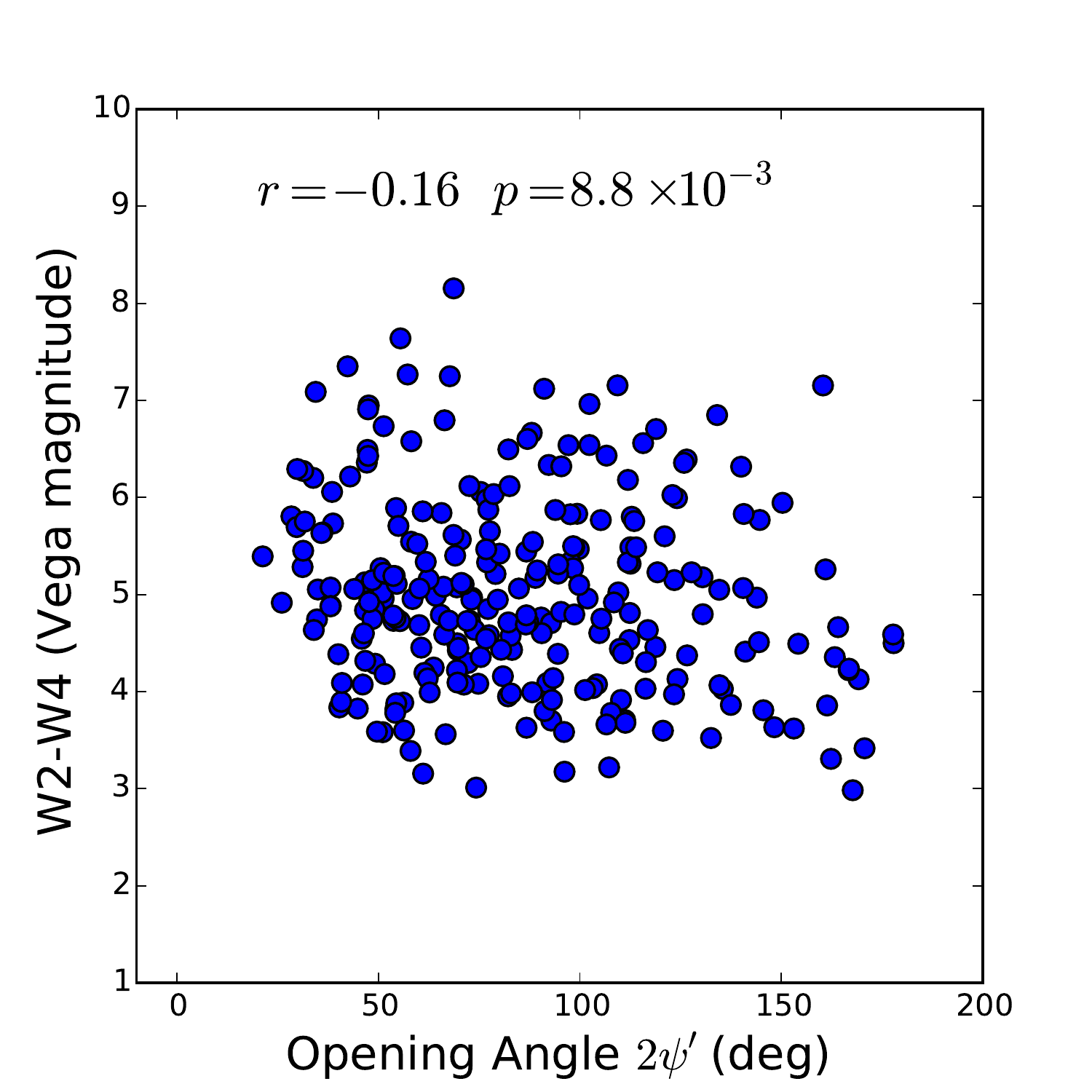}
\includegraphics[height=7.9cm,angle=0]{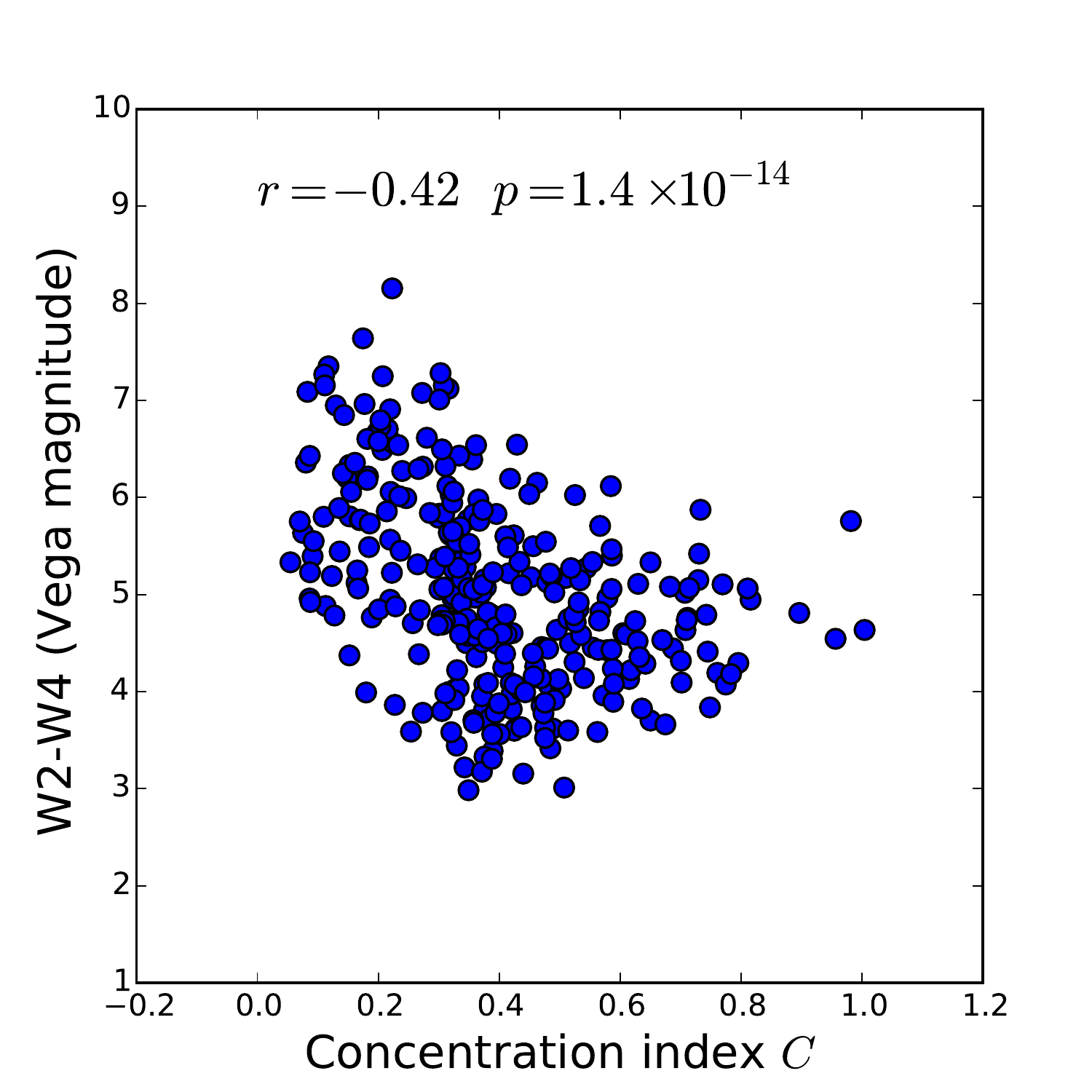}
\caption{Infrared color (W4 - W2) vs the parameters of morphology of AGN ionization region $A_2/A_0$, opening angle and concentration index 
$C$. The rank correlation coefficient $r$ and the probability of the null hypothesis $p$ are indicated in every panel.}
\label{fig7}
\end{figure*}

The correlations between the mid-IR color and the NLR morphological parameters ($A_2/A_0$, 2$\psi'$, and $C$) are shown in Fig~\ref{fig7}. 
All three of these relations have statistically significant correlations in a sense that they are all consistent with the expectations based on the unification model laid out above. 
The correlations of W2-W4 with the bicone strength $A_2/A_0$ (Pearson's $r = 0.34$,~$p$-value $= 1.3 \times 10^{-9}$) 
and the concentration $C$ ($r = -0.42$,~$p$-value $< 10^{-9}$) are stronger than the one with the opening 
angle 2$\psi'$ ($r = -0.16$,~$p$-value $= 8.8 \times 10^{-3}$). 
The mid-IR colors used here are based on original fluxes without subtracting the stellar continuum. Adopting the stellar continuum
subtracted fluxes does not change the results significantly (see Sec. \ref{sec:data:wise}).
So, we only present the results with no corrections. 
Our result is consistent with \citet{rose2015}, who found that a
larger AGN inclination angle corresponds to a redder W2-W4 color.

The existence of these correlations suggests that the morphology of the narrow-line region on kpc scale is connected to the dusty structure in the AGN vicinity on pc scales. The implications of these correlations are discussed in Sec. \ref{sec:discussion}.

\begin{table*}
\centering
\caption{Summary of the Spearman correlation coefficients $r$ between the NLR morphological parameters (Bi-polar prominence $A_2/A_0$,
the projected half opening angle~$\psi'$, the concentration index~$C$), infrared color (W4 - W2) and the ratio of the galaxies' semi-minor to semi-major axes ($b/a$). The value in the bracket is the $p-$value.}
\scriptsize
\label{table1}
\begin{tabular}{lccccccccccccccccccc}
\hline
      & $\psi'$ & $C$  &  W2-W4    & $b/a$ (disks) & $b/a$ (ellipticals) & $b/a$ (All) \\
\hline
$A_2/A_0$   & -0.55 ($9.5\times 10^{-26}$)&-0.3 ($7.3\times 10^{-8}$) & 0.34 ($1.3\times 10^{-9}$) &-0.31 ($4.2\times 10^{-6}$) &-0.31 ($6.6\times 10^{-3}$)&-0.30 ($1.3\times 10^{-7}$) \\
$\psi'$  & -       & 0.01 ($8.7\times 10^{-1}$) & -0.16 ($8.8\times 10^{-3}$) & 0.19 ($1.1\times 10^{-2}$)  & 0.29 ($3.1\times 10^{-2}$)& 0.20 ($1.5\times 10^{-3}$)    \\
 $C$  & -   & -     & -0.42 ($1.4\times 10^{-14}$) & 0.03 ($6.9\times 10^{-1}$) & 0.06 ($6.1\times 10^{-1}$)& -0.02 ($7.7\times 10^{-1}$)\\
 W2-W4   & -        &  -  &  -        &-0.18 ($8.6\times 10^{-3}$)  &-0.11 ($3.3\times 10^{-1}$) &-0.12 ($4.0\times 10^{-2}$)     \\
\hline
\label{t1}
\end{tabular}
\scriptsize
\end{table*}

\subsection{Alignment Between Narrow-line Region and Host Galaxy}
\label{sec:result:align}

In the previous section, we have found evidence suggesting that the shape of the narrow-line region may be linked to the mid-IR emitting dusty structure on pc-scales, in agreement with the unification model. 
But dust on larger scales, for example in the galactic disk, may also obscure the ionizing radiation and contribute to shaping the narrow-line region, in which case, we may expect correlations between the orientation of the ionization cones and the galaxy \citep{Lacy2007}. 
In this section, we investigate the links between the narrow-line regions and their host galaxies by correlating their position angles and comparing their morphological parameters. 

\begin{figure}
\center{}
\includegraphics[height=7.8cm,angle=0]{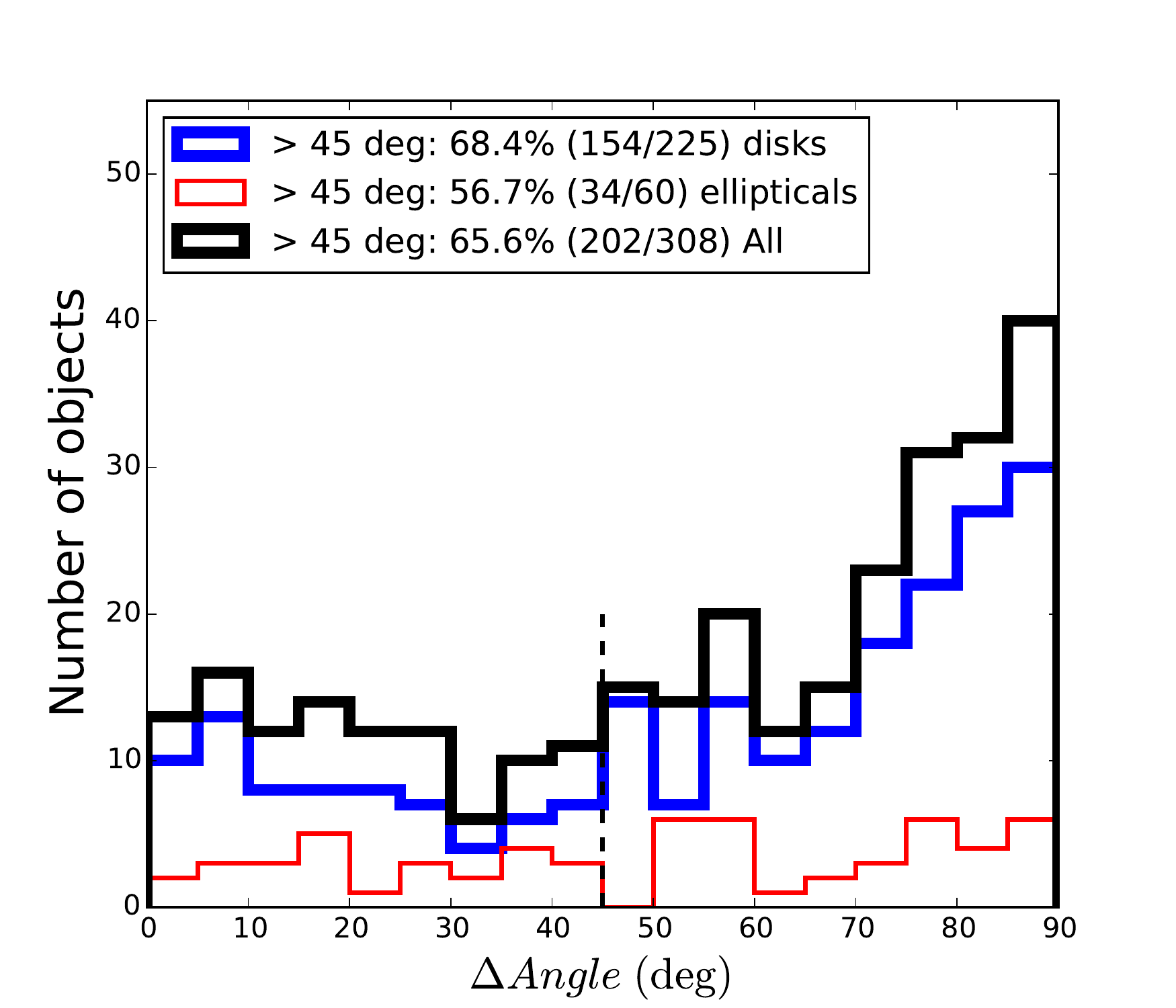}
\caption{Distribution of acute angles between the polar axis bicone and major axis of galaxies disk. 
Most of the value is close to 
90$^{o}$ which means the polar axis of bicone and major axis of galaxies disk are in the same direction.
For the whole sample, 202 out of 308 angles are larger than 45 deg. There are 154/225 and 34/60 angles larger than
45 deg for the subsample of disks and ellipticals respectively.
The probability $P(>45^o)$ is significant (binomial test, $p$-value $= 4.9\times 10^{-8},3.2\times 10^{-8}$ for the whole sample and
disks respectively) different from $P(>45^o) = 0.5$. 
There is also a significant difference between acute angles distribution and the uniform distribution 
(Kolmogorov--Smirnov test,~$p$-value$ < 10^{-9}$ for the whole sample,~$p$-value$ < 10^{-9}$ for disks). 
The dashed vertical line marks the location of 45 deg.}
\label{fig8}
\end{figure}

First, we look into the alignment between the narrow-line region and the host galaxy in the plane of the sky.  
The position angle of the ionization cone is determined with Fourier decomposition of the narrow-line region map (Sec.~\ref{subsec:bipolar}), and that of the host galaxy is from NASA--Sloan Atlas catalog (Sec.~\ref{sec:data:sdss}).
To determine their relative orientation, we measure the acute angle between the two position angles.
As shown in Fig.~\ref{fig8}, most of the acute angles between the bicones and the galaxies is closer to 90 degrees, i.e., the cones are preferentially perpendicular to the major-axis of the galaxy. 

To test the significance of the alignment, we compare the distribution of the acute angles with a uniform distribution, which would be the case if there were no alignment between the two components. The Kolmogorov--Smirnov test gives a statistically significant result ($r$ = 0.19,~$p$-value $< 10^{-9}$ for the whole sample,~$r$ = 0.22,~$p$-value $< 10^{-9}$ for disks), ruling out the case of random alignment. 
In addition, we divide the sample into two groups -- ones with acute angle larger than 45 deg (more perpendicular) and ones smaller than 45 degrees (more parallel). 
For the whole sample, 202 out of 308 angles are larger than 45 deg. There are 154/225 (68.4\%) and 34/60 (56.7\%) angles larger than
45 deg for the subsample of disks and ellipticals respectively.
The probability of $P(>45^o)$ is significant higher than 0.5 for the whole sample and disks (binomial test, $p$-value $= 4.9\times 10^{-8},3.2\times 10^{-8}$ for the whole sample, disks respectively). 
These results indicate that the polar-axis of the ionization cones and the major-axis of the galaxies disks are preferentially 
perpendicular to each other.

Second, additional tests can be made based on the inclination angles of the ionization cones, the dusty torus, and the galaxy. 
The galaxy's inclination is better defined with disky galaxies, for which the ratio between the semi-minor to semi-major axis ($b/a$) has been used as a proxy of inclination. 
We use the parameter \texttt{nsa\_sersic\_ba} from NASA--Sloan Atlas catalog (Sec. \ref{sec:data:sdss}) and expect this ratio to be lower when the galaxy is more inclined. 
For elliptical galaxies, although such a ratio is also measured, it is less clear what it means for the galaxy's orientation. 
Based on the unification model as discussed in Sec. \ref{sec:Measurement}, there are three morphological parameters of NLR that should be correlated with the inclination of the cones -- the strength of the bicone $A_2/A_0$, the opening angle 2$\psi'$ and the concentration $C$. The first one increases and the later two decrease with the inclination angle. 
For the dusty torus, its inclination is expected to the positively correlated with the mid-IR color W2-W4. 

\begin{figure*}
\center{}
\includegraphics[height=7.9cm,angle=0]{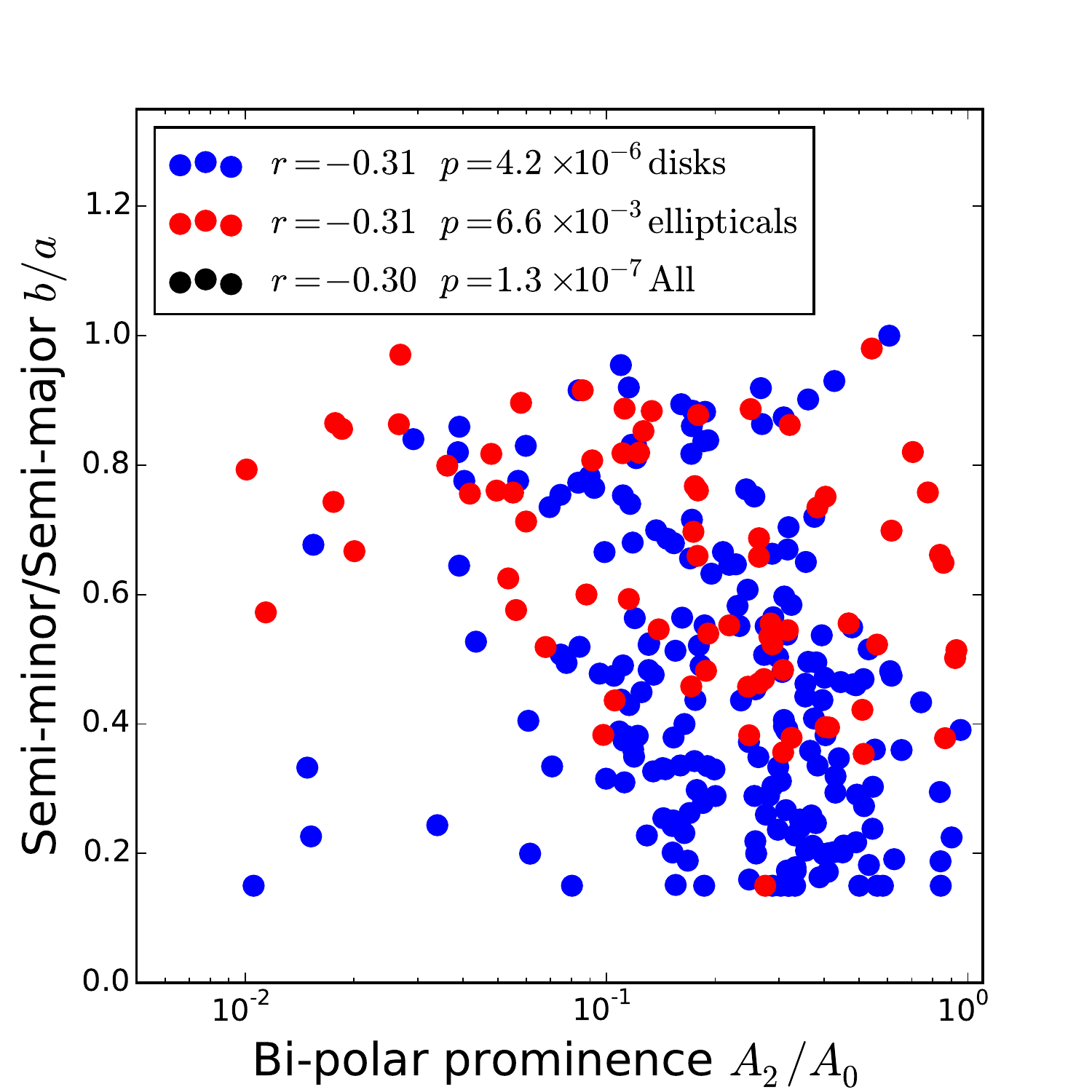}
\includegraphics[height=7.9cm,angle=0]{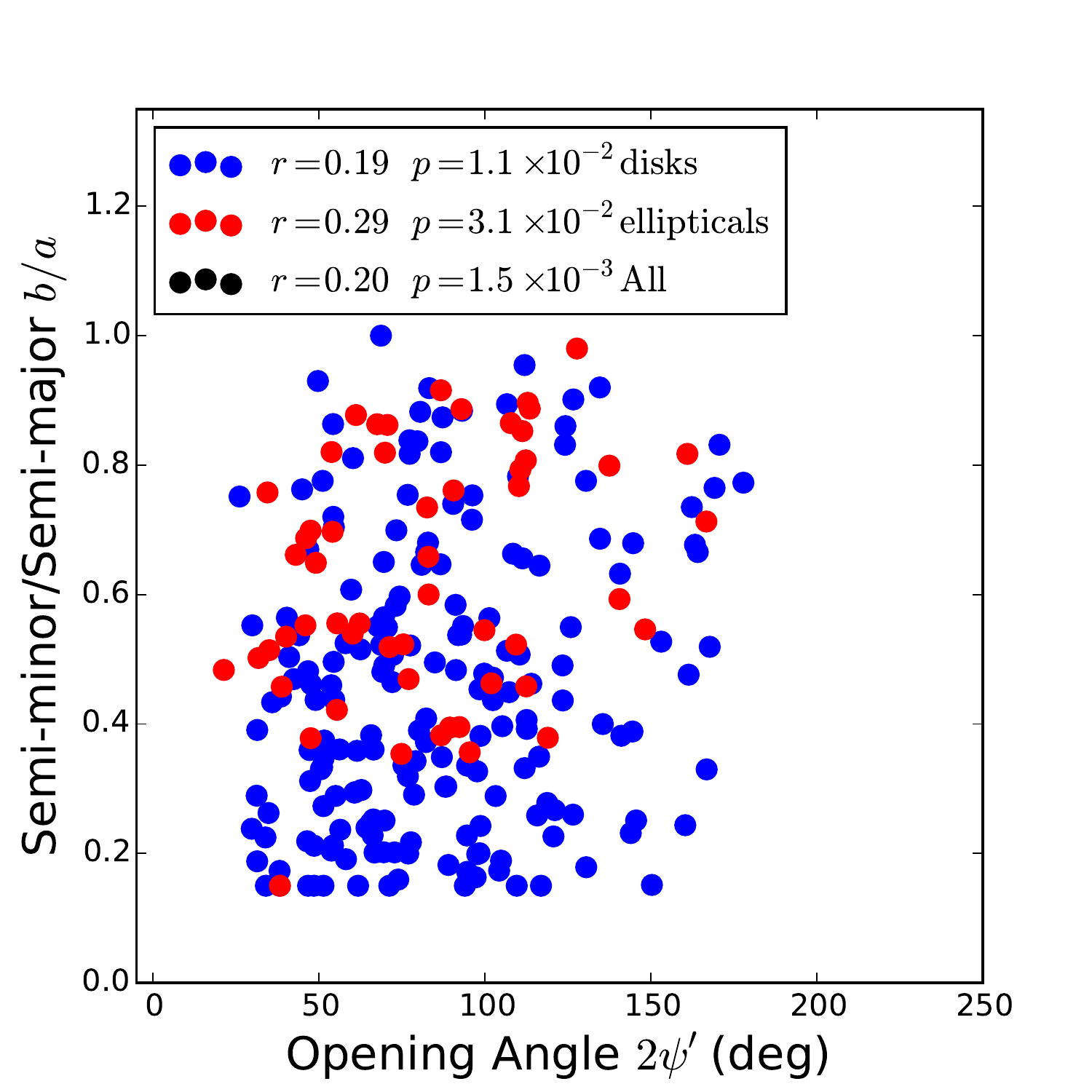}
\includegraphics[height=7.9cm,angle=0]{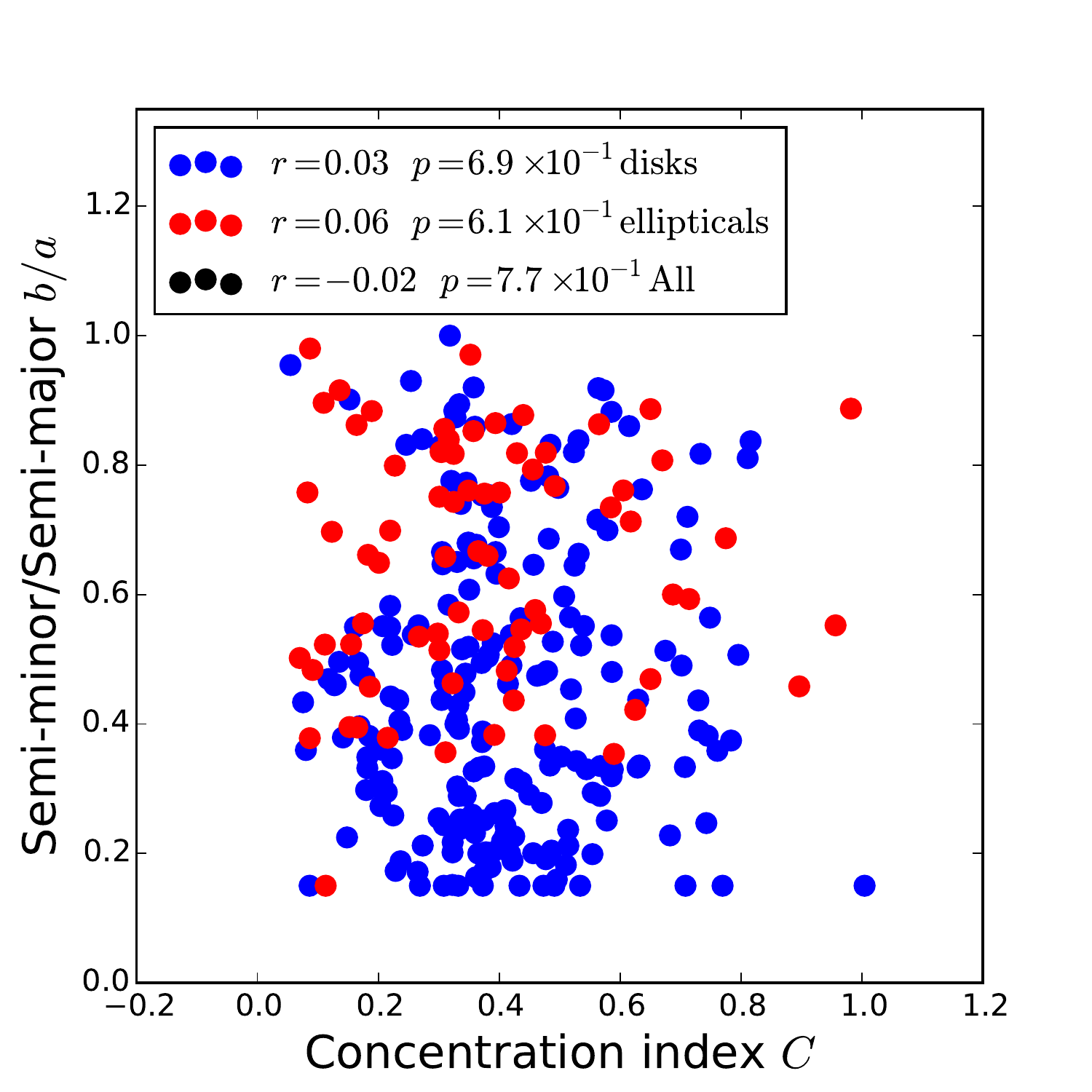}
\includegraphics[height=7.9cm,angle=0]{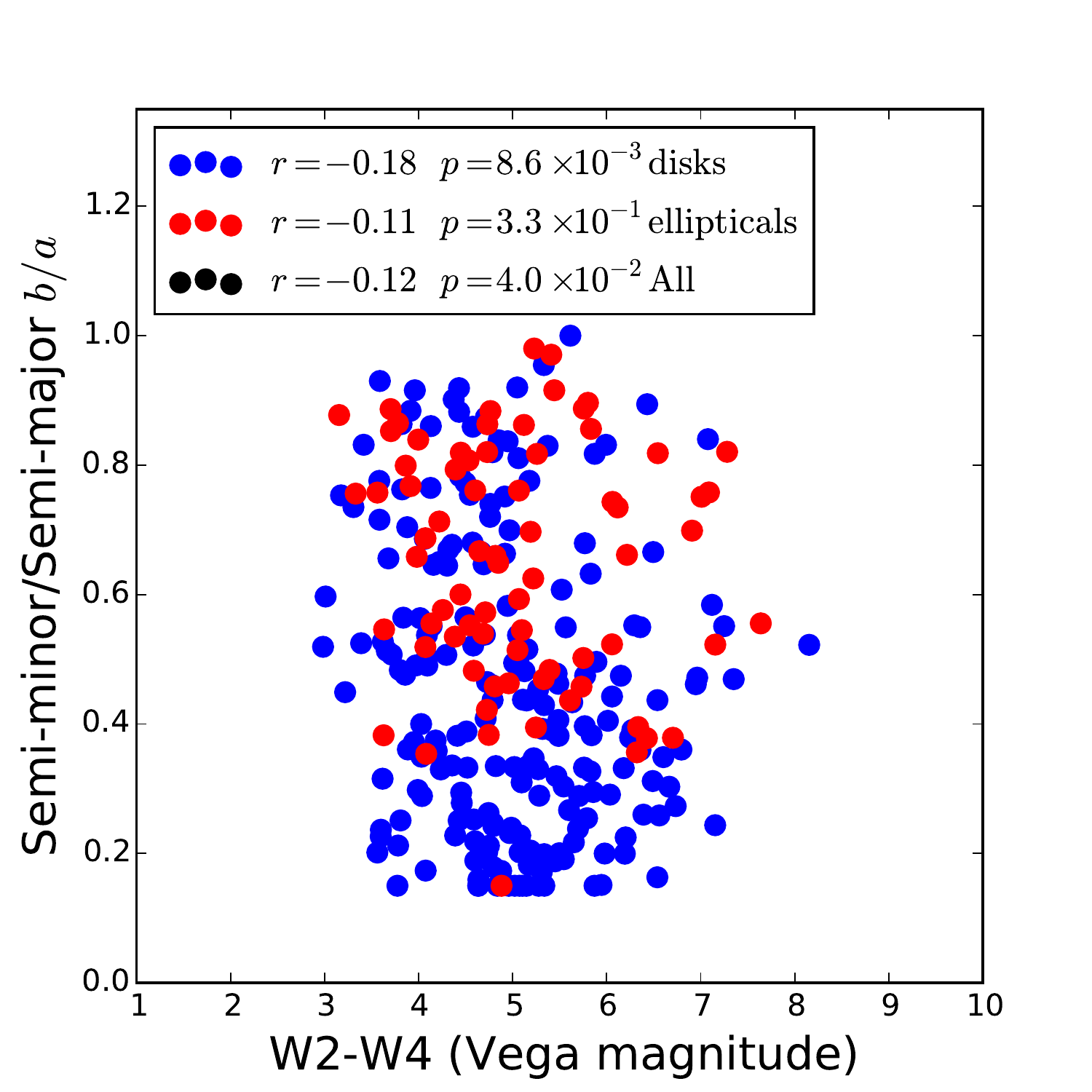}

\caption{The ratio of semi-minor to major axis of the galaxy vs the parameters of morphology of AGN ionization region $A_2/A_0$, opening angle, concentration index C and the infrared color (W4 - W2). Red points represent the ellipticals while the blue ones represent the disk galaxies. The rank correlation coefficient $r$ and the probability of the null hypothesis $p$ are indicated in every panel.
}
\label{fig9}
\end{figure*}

Fig.~\ref{fig9} shows the relations between the galaxy's semi-minor to semi-major ratio and the other inclination proxies. Their correlation coefficients are calculated based on the disk galaxies, elliptical galaxies, and the entire sample, respectively.
We find that, based on the whole sample, the ratio of galaxy's semi-minor to semi-major axis ($b/a$) is significantly anti-correlated with the prominence of the ionization cone $A_2/A_0$  ($r = -0.3$,~$p$-value $=  1.3\times 10^{-7}$), in agreement with our expectation that both depend on the inclination angle. 
The correlations between $b/a$ and the other parameters --- the cones' opening angle 2$\psi'$, concentration $C$, and the W2-W4 color --- are not significant.

\section{Discussion} \label{sec:discussion}

\subsection{Tests of Unification Model -- Type I / Type II ratio} \label{sec:discussion:ratio}

In the unification model, the opening angle of the ionization cones determine the number ratio between type I and type II AGN. 
As a test to the unification model, we compare our results to direct demographic studies of the type ratio. 

The opening angle measurements in our sample indicate a type II fraction of $\sim 60 - 70 \%$ at luminosities of \loiii=$10^{39.3} - 10^{41}$ \erg{}. 
Radio and mid-IR selected samples give type II fraction $\sim 50 - 60 \%$ that does not depend strongly on the AGN luminosity across a large range of luminosities \lbol$=10^{43-47}$ \erg{} \citep[][and references therein]{Lawrence2010}. 
Optically selected samples from SDSS suggest that the type II fraction decreases with the \OIII{} luminosity \citep{Simpson2005, Hao2005a}. 
For low luminosities AGN comparable to our sample (\loiii~$\sim 10^{40-41}$ \erg), \cite{Simpson2005} found that the Type II fraction is around $80-90\%$, which is higher than what we find here.
At high luminosities (\loiii~$\sim 10^{42-43}$ \erg), \citet{Simpson2005, Reyes2008} find a lower type II ratio of $50-70\%$. 
In X-ray selected samples, the optically defined type II AGN fraction is $\sim 60-90\%$ in the low luminosity range of $L_{\rm{2-10~keV}} = 10^{42.5-43.5}$ that is comparable to our sample \citep{Merloni2014}. 

Although there are discrepancies in the type II AGN fraction on the order of 10-20\% between our results and other studies, variation of this level is seen among different demographic studies. 
These variations are likely due to the selection biases associated with different AGN selection methods, different redshift coverage, and different classification criteria for type I versus type II AGN. 
We conclude that our results are broadly in agreement with demographic studies and that there is no strong evidence conflicting with the unification model. 

With Hubble Space Telescope (HST) imaging, \citet{Obied2016} constrain the intrinsic half opening angle of the AGN scattered light cone to be $27^\circ \pm 9^\circ$ (mean and standard deviation) among luminous type II AGN, which is smaller than our result implying a higher type II fraction. It is unclear what causes the discrepancy between \citet{Obied2016} and our result. \citet{Obied2016} points out that their results are inconsistent with demographic AGN type ratios at comparable luminosity and redshift range and speculate that the type I fraction could be increased by allowing illuminating light to escape through a porous torus. We do not see strong evidence for such component in our sample.

The intrinsic opening angle and the type I/II ratio inferred in our study is subject to a few sources of uncertainties. First of all, the discrete binning along the azimuthal angle and less-distinct cone boundaries introduce uncertainties of the order $\sim 10^\circ$ in the measured projected opening angle, corresponding to $10\%$ uncertainty in the type II fraction. 
Second, the morphology of the narrow-line regions may also be affected by factors in the host galaxy, including the gas distribution, dust extinction, or contamination of the emission lines from star-forming regions, although these uncertainties are harder to quantify. 
Third, given the observed projected opening angles, the inferred ratio also depends on the assumed function form of the intrinsic opening angle distribution. Our assumed Gaussian and Beta profiles provide satisfactory fit to the data. But in the extreme case if there is a missing population of AGN with no obscuration at all, such a population can lower the type II AGN fraction but would not be reflected in our results. 
In another extreme, we would miss the completely enshrouded AGN which do not have photo-ionized emission-line regions.

In addition, there is a small fraction (12\%, 38/308) of our type II AGN sample showing halo-like NLR morphology, which is rather surprising because such a morphology is expected in type I AGN in the unification model. This may be because the cones exist but they are just too small to be spatially resolved.  Or alternatively, this may indicate a dust configuration more complex than the torus model. 
We may indeed be looking into the ionization cone, but the direct line-of-sight to the nucleus is blocked by individual clouds within the cone. In this case, the correspondence between the cone opening angle and the type ratio would be more complicated.

\subsection{Tests of Unification Model -- Source of Obscuration}
The AGN photo-ionized emission regions reflect the illumination geometry, which in turn is shaped by the obscuration. 
Previous studies suggest that dust on either circumnuclear or galactic scales can obscure the AGN radiation,  \cite[e.g.,][]{Lacy2007}. 

Indeed, we see some evidence in favor of the obscuration occurring on circumnuclear scales. We see that as the cones become narrower or the prominence of the cone $A_2/A_0$ becomes larger, the IR colors become redder.
The \loiii~ for all objects in our sample are $\sim10^{39.3}$ to $10^{41}$\erg.
Applying the relationship between ~\loiii~ and ~\lbol~ for Type 2 AGN \citep{Liu2009}, we find the typical ~\lbol~ to be $10^{42.4}$--$10^{44.1}$\erg. Adopting the calculation in \citep{Barvainis1987, Elsner1987, Jiang2017}, the IR emission at W2 to W4 band (4.6--22$\mu m,\rm T \sim 500K$) is produced on the scale $R\sim~0.02-1\rm~pc$. Therefore, if there is a relationship between photo-ionized regions and the warm IR, the obscuration and the illumination must be established on these small scales.


In this work, we find that the higher concentration of the AGN NLR emission corresponds to hotter dust, consistent with the expectation that type I AGN or AGN with small inclination have higher NLR concentration \citep{Mulchaey1996,Schmitt2003}.
In addition, \cite{Schmitt2003} did not find type I to have smaller NLR than type II, but they have compared their morphology: the morphology of ionized regions of Seyfert I are more round shaped and that of Seyfert II are more cone like. 
These findings, together with ours, all suggest that the apparent concentration of the NLR is largely determined by the inclination angle. 

We also see that the orientation of the AGN ionization cones is related to the position angle (major axis) of 
the galaxy. In disk galaxies, the ratio of galaxy's semi-minor to semi-major axis ($b/a$) is significantly anti-correlated 
with the prominence of the ionization cone $A_2/A_0$.
Our measurements show that the ionization region cones are preferentially orthogonal to the major axis of the galaxy.
This could be because the orientation of the obscuring material is correlated with the orientation of the galaxy disk. 
Despite a vast difference of physical scales, there are some theoretical models of AGN activity in which there is a relationship between the accretion disk and the galactic disk \citep{Shlosman1989, Hopkins2009, Hopkins2010}.
Given the fueling of galactic gas onto the AGN, in these models, one may expect a certain degree of alignment between the AGN obsuration and the host galaxy disk. 
However, \cite{Pjanka2017} find no evidence for megamaser disks preferentially aligning with the galactic disks casting doubt on the physical connection between AGN obscuration and the galactic orientation. 
Alternatively, the correlation between the NLR and disk orientation could have something to do with the extinction: in the disk, we would have trouble seeing photo-ionized gas beyond a distance of kiloparsecs \citep{Lacy2007}.
From our evidence of the connections between ionization cone and IR color and the galaxy disk,
we conclude that both circumnuclear obscuration and the host galaxy disk play a part in the morphology of AGN NLRs.

A subset of our AGN show ring-like morphologies of the NLR with a hole in the center (top panel of Fig.~\ref{fig4}). This morphology is not well-captured by the unification model. One possible explanation is that the central engine is in a period of inactivity at a small inclination angle.
Another possible explanation is that the central engine is obscured by a dusty cloud or dominated by a nuclear star-forming regions.

\section{Conclusion}
\label{sec:conclusion}

The geometry of the AGN obscuration remains an active topic in the discussion of AGN unification model. 
In this paper, we approach this question with statistical studies of the narrow-line region morphology. 
Our sample consists of 308 type II AGN candidates selected from the MaNGA IFU survey \citep{Wylezalek2018}, which offer spatially resolved 2D narrow-line region maps. 

To identify the bi-conical shape of the ionized region, we decompose the azimuthal dependence of the narrow-line region maps into Fourier series. 
We find that the power of the $m = 2$ mode ($A_2/A_0$) signals the presence of the bicones. 
The phase of the $m=2$ mode corresponds to the position angle of the cones. 
The opening angles of the cones are characterized by the full-width-half-maximum (FWHM) of the narrow-line region azimuthal profile. 
81 percent of the AGN are found to have bi-conical or bi-polar narrow-line regions. 
The results of automated morphological measurements are robust and are in good agreement with visual inspection. 

We use the distribution of the measured opening angles to infer the intrinsic opening angles of the ionization cones assuming random projection. We find that the data are inconsistent with a single intrinsic opening angle. 
To account for the spread, we model the distribution of the intrinsic opening angle with a Gaussian and a Beta function. 
With MCMC fitting, we constrain the mean and the standard deviation of the intrinsic opening angles in our sample to be 85.2 -- 97.6 and 39.2 -- 44.2 deg. 
Our result suggests that the number ratio of type I and type II AGN should be $\approx~1 : 1.6 - 2.3$, or a type II AGN fraction of 60\% to 70\%, which is
consistent with the observed number ratios of type I to type II at low luminosities.

We find that the NLR morphological parameters, i.e., bicone strength $A_2/A_0$, opening angle and concentration index, are significantly correlated with the WISE mid-IR color W2-W4, which reflects the circumnuclear dust geometry. 
The correlation coefficients between W2-W4 colors and the morphological parameters are $r = 0.35$,~$p$-value $< 10^{-9}$, $r = -0.42$,
~$p$-value $ < 10^{-9}$, and $r = -0.16$,~$p$-value $=  8.4 \times 10^{-3}$, respectively, for the bicone strength, opening angle, and concentration index. 
This indicates that AGN with redder mid-IR colors (lower temperature of dust or more edge-on dusty torus) corresponds to a more prominent, narrower, and less centrally concentrated bicone morphology. 
These results are in agreement with the expectations that the orientation of ionization cones is related that of the toroidal circumnuclear dusty structure. 

Furthermore, we find significant evidence (binomial test, $p$-value $= 4.9\times 10^{-8}$) that the major-axis of the galaxy disk and the AGN ionization cones are preferentially orthogonal to each other. 
In addition, we find correlations between indicators of galaxies' and ionization cones' inclinations. 
The ratios of the galaxies' semi-minor to semi-major axes ($b/a$) are anti-correlated with the prominence of the ionization cones $A_2/A_0$ and positively correlated with the cone opening angles. 
These relationships between NLRs and the host galaxies could arise when the orientation of the obscuring material is correlated with the orientation of the galaxy disk. Alternatively, this
could be due to galactic disk extinction contributing to shaping the ionizing radiation or that the emission lines are obscured on galactic scales.

It has long been debated whether AGN obscuration is due to circumnuclear material, to the galactic disk, or both. We find evidence for both processes in our study. In future studies of AGN demographics, it will be important to quantify the relative importance of these two contributions as a function of AGN luminosity and host galaxy types. 

\section{ACKNOWLEDGMENTS}
We would like to thank Guilin Liu for useful discussion.
Zhicheng He is supported by China Scholarship Council (CSC, NO. 201706340030) during his stay at the Johns Hopkins University.
Zhicheng He acknowledges the grant from the National Natural Science Foundation of China (No. 11673020 and NO. 11421303) and the Ministry of Science and Technology of China (National Key Program for Science and Technology Research and Development, NO. 2016YFA0400700).
Nadia L. Zakamska acknowledges support by the Johns Hopkins University provided through the Catalyst award. 
Rog\'erio Riffel thanks to CNPq and FAPERGS for partial funding this project.
Rogemar A. Riffel acknowledges support from FAPERGS (projects NO 16/2551-0000 and 17/2551-0001) and CNPq (project NO 303373/2016-4).

Funding for the Sloan Digital Sky Survey IV has been provided by the Alfred P. Sloan Foundation, the U.S. Department of Energy Office of Science, and the Participating Institutions. SDSS-IV acknowledges
support and resources from the Center for High-Performance Computing at
the University of Utah. The SDSS web site is www.sdss.org.

SDSS-IV is managed by the Astrophysical Research Consortium for the 
Participating Institutions of the SDSS Collaboration including the 
Brazilian Participation Group, the Carnegie Institution for Science, 
Carnegie Mellon University, the Chilean Participation Group, the French Participation Group, Harvard-Smithsonian Center for Astrophysics, 
Instituto de Astrof\'isica de Canarias, The Johns Hopkins University, 
Kavli Institute for the Physics and Mathematics of the Universe (IPMU) / 
University of Tokyo, Lawrence Berkeley National Laboratory, 
Leibniz Institut f\"ur Astrophysik Potsdam (AIP),  
Max-Planck-Institut f\"ur Astronomie (MPIA Heidelberg), 
Max-Planck-Institut f\"ur Astrophysik (MPA Garching), 
Max-Planck-Institut f\"ur Extraterrestrische Physik (MPE), 
National Astronomical Observatories of China, New Mexico State University, 
New York University, University of Notre Dame, 
Observat\'ario Nacional / MCTI, The Ohio State University, 
Pennsylvania State University, Shanghai Astronomical Observatory, 
United Kingdom Participation Group,
Universidad Nacional Aut\'onoma de M\'exico, University of Arizona, 
University of Colorado Boulder, University of Oxford, University of Portsmouth, 
University of Utah, University of Virginia, University of Washington, University of Wisconsin, 
Vanderbilt University, and Yale University.

This publication makes use of data products from the Wide-field Infrared Survey Explorer, which is a joint project of the University of California, Los Angeles, and the Jet Propulsion Laboratory/California Institute of Technology, funded by the National Aeronautics and Space Administration. 

\bibliography{local}

\begin{thebibliography}{}
\makeatletter
\relax
\def\mn@urlcharsother{\let\do\@makeother \do\$\do\&\do\#\do\^\do\_\do\%\do\~}
\def\mn@doi{\begingroup\mn@urlcharsother \@ifnextchar [ {\mn@doi@}
  {\mn@doi@[]}}
\def\mn@doi@[#1]#2{\def\@tempa{#1}\ifx\@tempa\@empty \href
  {http://dx.doi.org/#2} {doi:#2}\else \href {http://dx.doi.org/#2} {#1}\fi
  \endgroup}
\def\mn@eprint#1#2{\mn@eprint@#1:#2::\@nil}
\def\mn@eprint@arXiv#1{\href {http://arxiv.org/abs/#1} {{\tt arXiv:#1}}}
\def\mn@eprint@dblp#1{\href {http://dblp.uni-trier.de/rec/bibtex/#1.xml}
  {dblp:#1}}
\def\mn@eprint@#1:#2:#3:#4\@nil{\def\@tempa {#1}\def\@tempb {#2}\def\@tempc
  {#3}\ifx \@tempc \@empty \let \@tempc \@tempb \let \@tempb \@tempa \fi \ifx
  \@tempb \@empty \def\@tempb {arXiv}\fi \@ifundefined
  {mn@eprint@\@tempb}{\@tempb:\@tempc}{\expandafter \expandafter \csname
  mn@eprint@\@tempb\endcsname \expandafter{\@tempc}}}

\bibitem[\protect\citeauthoryear{Antonucci}{Antonucci}{1993}]{Antonucci1993}
Antonucci R.,  1993, \mn@doi [Annu. Rev. Astron. Astrophys.]
  {10.1146/annurev.aa.31.090193.002353}, 31, 473

\bibitem[\protect\citeauthoryear{Antonucci \& Miller}{Antonucci \&
  Miller}{1985}]{Antonucci1985}
Antonucci R. R.~J.,  Miller J.~S.,  1985, \mn@doi [Astrophys. J.]
  {10.1086/163559}, 297, 621

\bibitem[\protect\citeauthoryear{{Audibert}, {Riffel}, {Sales}, {Pastoriza}  \&
  {Ruschel-Dutra}}{{Audibert} et~al.}{2017}]{Audibert2017}
{Audibert} A.,  {Riffel} R.,  {Sales} D.~A.,  {Pastoriza} M.~G.,
  {Ruschel-Dutra} D.,  2017, \mn@doi [\mnras] {10.1093/mnras/stw2477}, \href
  {http://adsabs.harvard.edu/abs/2017MNRAS.464.2139A} {464, 2139}

\bibitem[\protect\citeauthoryear{Baldwin \& Terlevich}{Baldwin \&
  Terlevich}{1981}]{Baldwin1981}
Baldwin J~A P. M.~M.,  Terlevich R.,  1981, \mn@doi [Publ. Astron. Soc.
  Pacific] {10.1086/130766}, 93, 5

\bibitem[\protect\citeauthoryear{Barrera-Ballesteros
  et~al.,}{Barrera-Ballesteros et~al.}{2016}]{Barrera-Ballesteros2016}
Barrera-Ballesteros J.~K.,  et~al., 2016, \mn@doi [Monthly Notices of the Royal
  Astronomical Society, Volume 463, Issue 3, p.2513-2522]
  {10.1093/mnras/stw1984}, 463, 2513

\bibitem[\protect\citeauthoryear{Barvainis \& Richard}{Barvainis \&
  Richard}{1987}]{Barvainis1987}
Barvainis R.,  Richard 1987, \mn@doi [The Astrophysical Journal]
  {10.1086/165571}, 320, 537

\bibitem[\protect\citeauthoryear{Bershady, Jangren  \& Conselice}{Bershady
  et~al.}{2000}]{Bershady2000}
Bershady M.~A.,  Jangren A.,   Conselice C.~J.,  2000, \mn@doi [The
  Astronomical Journal, Volume 119, Issue 6, pp. 2645-2663.] {10.1086/301386},
  119, 2645

\bibitem[\protect\citeauthoryear{Blanton, Kazin, Muna, Weaver  \&
  Price-Whelan}{Blanton et~al.}{2011}]{Blanton2011}
Blanton M.~R.,  Kazin E.,  Muna D.,  Weaver B.~A.,   Price-Whelan A.,  2011,
  \mn@doi [The Astronomical Journal, Volume 142, Issue 1, article id. 31, 14
  pp. (2011).] {10.1088/0004-6256/142/1/31}, 142

\bibitem[\protect\citeauthoryear{Blanton et~al.,}{Blanton
  et~al.}{2017}]{Blanton2017}
Blanton M.~R.,  et~al., 2017, \mn@doi [The Astronomical Journal, Volume 154,
  Issue 1, article id. 28, 35 pp. (2017).] {10.3847/1538-3881/aa7567}, 154

\bibitem[\protect\citeauthoryear{Bundy et~al.,}{Bundy et~al.}{2014}]{Bundy2014}
Bundy K.,  et~al., 2014, \mn@doi [Astrophys. J.] {10.1088/0004-637X/798/1/7},
  798, 7

\bibitem[\protect\citeauthoryear{Doi et~al.,}{Doi et~al.}{2010}]{Doi2010}
Doi M.,  et~al., 2010, \mn@doi [The Astronomical Journal, Volume 139, Issue 4,
  pp. 1628-1648 (2010).] {10.1088/0004-6256/139/4/1628}, 139, 1628

\bibitem[\protect\citeauthoryear{Drory et~al.,}{Drory et~al.}{2014}]{Drory2014}
Drory N.,  et~al., 2014, \mn@doi [The Astronomical Journal, Volume 149, Issue
  2, article id. 77, 24 pp. (2015).] {10.1088/0004-6256/149/2/77}, 149

\bibitem[\protect\citeauthoryear{Elitzur}{Elitzur}{2012}]{Elitzur2012}
Elitzur M.,  2012, \mn@doi [Astrophys. J.] {10.1088/2041-8205/747/2/L33}, 747,
  L33

\bibitem[\protect\citeauthoryear{Elsner, Shibazaki  \& Weisskopf}{Elsner
  et~al.}{1987}]{Elsner1987}
Elsner R.~F.,  Shibazaki N.,   Weisskopf M.~C.,  1987, \mn@doi [The
  Astrophysical Journal] {10.1086/165570}, 320, 527

\bibitem[\protect\citeauthoryear{Evans, Ford, Kinney, Antonucci, Armus  \&
  Caganoff}{Evans et~al.}{1991}]{Evans1991}
Evans I.~N.,  Ford H.~C.,  Kinney A.~L.,  Antonucci R. R.~J.,  Armus L.,
  Caganoff S.,  1991, \mn@doi [Astrophys. J.] {10.1086/185951}, 369, L27

\bibitem[\protect\citeauthoryear{Ezhikode, Gandhi, Done, Ward, Dewangan, Misra
  \& Philip}{Ezhikode et~al.}{2017}]{Ezhikode2017}
Ezhikode S.~H.,  Gandhi P.,  Done C.,  Ward M.,  Dewangan G.~C.,  Misra R.,
  Philip N.~S.,  2017, \mn@doi [Mon. Not. R. Astron. Soc.]
  {10.1093/mnras/stx2160}, 472, 3492

\bibitem[\protect\citeauthoryear{Fischer, Crenshaw, Kraemer  \&
  Schmitt}{Fischer et~al.}{2013}]{Fischer2013}
Fischer T.~C.,  Crenshaw D.~M.,  Kraemer S.~B.,   Schmitt H.~R.,  2013, \mn@doi
  [Astrophys. J. Suppl. Ser.] {10.1088/0067-0049/209/1/1}, 209, 1

\bibitem[\protect\citeauthoryear{Fischer, Crenshaw, Kraemer, Schmitt  \&
  Turner}{Fischer et~al.}{2014}]{Fischer2014}
Fischer T.~C.,  Crenshaw D.~M.,  Kraemer S.~B.,  Schmitt H.~R.,   Turner T.~J.,
   2014, \mn@doi [Astrophys. J.] {10.1088/0004-637X/785/1/25}, 785, 25

\bibitem[\protect\citeauthoryear{Garc{\'{i}}a-Gonz{\'{a}}lez
  et~al.,}{Garc{\'{i}}a-Gonz{\'{a}}lez et~al.}{2017}]{Garcia-Gonzalez2017}
Garc{\'{i}}a-Gonz{\'{a}}lez J.,  et~al., 2017, \mn@doi [Mon. Not. R. Astron.
  Soc.] {10.1093/mnras/stx1361}, 470, 2578

\bibitem[\protect\citeauthoryear{Goulding, Alexander, Bauer, Forman, Hickox,
  Jones, Mullaney  \& Trichas}{Goulding et~al.}{2012}]{Goulding2012}
Goulding A.~D.,  Alexander D.~M.,  Bauer F.~E.,  Forman W.~R.,  Hickox R.~C.,
  Jones C.,  Mullaney J.~R.,   Trichas M.,  2012, \mn@doi [Astrophys. J.]
  {10.1088/0004-637X/755/1/5}, 755, 5

\bibitem[\protect\citeauthoryear{Gunn et~al.,}{Gunn et~al.}{2006}]{Gunn2006}
Gunn J.~E.,  et~al., 2006, \mn@doi [The Astronomical Journal, Volume 131, Issue
  4, pp. 2332-2359.] {10.1086/500975}, 131, 2332

\bibitem[\protect\citeauthoryear{Hao et~al.,}{Hao et~al.}{2005}]{Hao2005a}
Hao L.,  et~al., 2005, \mn@doi [Astron. J.] {10.1086/428486}, 129, 1795

\bibitem[\protect\citeauthoryear{Hoenig, Kishimoto, Gandhi, Smette, Asmus,
  Duschl, Polletta  \& Weigelt}{Hoenig et~al.}{2010}]{Hoenig2010}
Hoenig S.~F.,  Kishimoto M.,  Gandhi P.,  Smette A.,  Asmus D.,  Duschl W.,
  Polletta M.,   Weigelt G.,  2010, \mn@doi [Astronomy and Astrophysics, Volume
  515, id.A23, 22 pp.] {10.1051/0004-6361/200913742}, 515

\bibitem[\protect\citeauthoryear{Hopkins \& Quataert}{Hopkins \&
  Quataert}{2009}]{Hopkins2009}
Hopkins P.~F.,  Quataert E.,  2009, \mn@doi [Monthly Notices of the Royal
  Astronomical Society, Volume 407, Issue 3, pp. 1529-1564.]
  {10.1111/j.1365-2966.2010.17064.x}, 407, 1529

\bibitem[\protect\citeauthoryear{Hopkins \& Quataert}{Hopkins \&
  Quataert}{2010}]{Hopkins2010}
Hopkins P.~F.,  Quataert E.,  2010, \mn@doi [Monthly Notices of the Royal
  Astronomical Society, Volume 415, Issue 2, pp. 1027-1050.]
  {10.1111/j.1365-2966.2011.18542.x}, 415, 1027

\bibitem[\protect\citeauthoryear{Ichikawa et~al.,}{Ichikawa
  et~al.}{2015}]{Ichikawa2015}
Ichikawa K.,  et~al., 2015, \mn@doi [Astrophys. J.]
  {10.1088/0004-637X/803/2/57}, 803, 57

\bibitem[\protect\citeauthoryear{Jarrett et~al.,}{Jarrett
  et~al.}{2011}]{Jarrett2011}
Jarrett T.~H.,  et~al., 2011, \mn@doi [Astrophys. J.]
  {10.1088/0004-637X/735/2/112}, 735, 112

\bibitem[\protect\citeauthoryear{Jiang et~al.,}{Jiang et~al.}{2017}]{Jiang2017}
Jiang N.,  et~al., 2017, \mn@doi [The Astrophysical Journal, Volume 850, Issue
  1, article id. 63, 7 pp. (2017).] {10.3847/1538-4357/aa93f5}, 850

\bibitem[\protect\citeauthoryear{Lacy, Sajina, Petric, Seymour, Canalizo,
  Ridgway, Armus  \& Storrie-Lombardi}{Lacy et~al.}{2007}]{Lacy2007}
Lacy M.,  Sajina A.,  Petric A.~O.,  Seymour N.,  Canalizo G.,  Ridgway S.~E.,
  Armus L.,   Storrie-Lombardi L.~J.,  2007, \mn@doi [The Astrophysical
  Journal, Volume 669, Issue 2, pp. L61-L64.] {10.1086/523851}, 669, L61

\bibitem[\protect\citeauthoryear{Lagos, Padilla, Strauss, Cora  \& Hao}{Lagos
  et~al.}{2011}]{Lagos2011}
Lagos C. d.~P.,  Padilla N.~D.,  Strauss M.~A.,  Cora S.~A.,   Hao L.,  2011,
  \mn@doi [Mon. Not. R. Astron. Soc.] {10.1111/j.1365-2966.2011.18531.x}, 414,
  2148

\bibitem[\protect\citeauthoryear{Law et~al.,}{Law et~al.}{2015}]{Law2015}
Law D.~R.,  et~al., 2015, \mn@doi [The Astronomical Journal, Volume 150, Issue
  1, article id. 19, 17 pp. (2015).] {10.1088/0004-6256/150/1/19}, 150

\bibitem[\protect\citeauthoryear{Lawrence \& Elvis}{Lawrence \&
  Elvis}{2010}]{Lawrence2010}
Lawrence A.,  Elvis M.,  2010, \mn@doi [The Astrophysical Journal]
  {10.1088/0004-637X/714/1/561}, 714, 561

\bibitem[\protect\citeauthoryear{{Liu}, {Zakamska}, {Greene}, {Strauss},
  {Krolik}  \& {Heckman}}{{Liu} et~al.}{2009}]{Liu2009}
{Liu} X.,  {Zakamska} N.~L.,  {Greene} J.~E.,  {Strauss} M.~A.,  {Krolik}
  J.~H.,   {Heckman} T.~M.,  2009, \mn@doi [\apj]
  {10.1088/0004-637X/702/2/1098}, \href
  {http://adsabs.harvard.edu/abs/2009ApJ...702.1098L} {702, 1098}

\bibitem[\protect\citeauthoryear{Lupton, Gunn  \& Szalay}{Lupton
  et~al.}{1999}]{Lupton1999}
Lupton R.,  Gunn J.,   Szalay A.,  1999, \mn@doi [The Astronomical Journal,
  Volume 118, Issue 3, pp. 1406-1410.] {10.1086/301004}, 118, 1406

\bibitem[\protect\citeauthoryear{Maiolino, Shemmer, Imanishi, Netzer, Oliva,
  Lutz  \& Sturm}{Maiolino et~al.}{2007}]{Maiolino2007}
Maiolino R.,  Shemmer O.,  Imanishi M.,  Netzer H.,  Oliva E.,  Lutz D.,
  Sturm E.,  2007, \mn@doi [Astron. Astrophys.] {10.1051/0004-6361:20077252},
  468, 979

\bibitem[\protect\citeauthoryear{Mateos et~al.,}{Mateos
  et~al.}{2016}]{Mateos2016}
Mateos S.,  et~al., 2016, \mn@doi [Astrophys. J.]
  {10.3847/0004-637X/819/2/166}, 819, 166

\bibitem[\protect\citeauthoryear{Merloni et~al.,}{Merloni
  et~al.}{2014}]{Merloni2014}
Merloni A.,  et~al., 2014, \mn@doi [Mon. Not. R. Astron. Soc.]
  {10.1093/mnras/stt2149}, 437, 3550

\bibitem[\protect\citeauthoryear{Mulchaey, Wilson  \& Tsvetanov}{Mulchaey
  et~al.}{1996a}]{Mulchaey1996}
Mulchaey J.~S.,  Wilson A.~S.,   Tsvetanov Z.,  1996a, \mn@doi [Astrophys. J.
  Suppl. Ser.] {10.1086/192261}, 102, 309

\bibitem[\protect\citeauthoryear{Mulchaey, Wilson  \& Tsvetanov}{Mulchaey
  et~al.}{1996b}]{Mulchaey1996a}
Mulchaey J.~S.,  Wilson A.~S.,   Tsvetanov Z.,  1996b, \mn@doi [Astrophys. J.]
  {10.1086/177595}, 467, 197

\bibitem[\protect\citeauthoryear{M{\"{u}}ller-S{\'{a}}nchez, Prieto, Hicks,
  Vives-Arias, Davies, Malkan, Tacconi  \& Genzel}{M{\"{u}}ller-S{\'{a}}nchez
  et~al.}{2011}]{Muller-Sanchez2011}
M{\"{u}}ller-S{\'{a}}nchez F.,  Prieto M.~A.,  Hicks E. K.~S.,  Vives-Arias H.,
   Davies R.~I.,  Malkan M.,  Tacconi L.~J.,   Genzel R.,  2011, \mn@doi
  [Astrophys. J.] {10.1088/0004-637X/739/2/69}, 739, 69

\bibitem[\protect\citeauthoryear{Netzer}{Netzer}{2015}]{Netzer2015}
Netzer H.,  2015, \mn@doi [Annu. Rev. Astron. Astrophys.]
  {10.1146/annurev-astro-082214-122302}, 53, 365

\bibitem[\protect\citeauthoryear{Obied, Zakamska, Wylezalek  \& Liu}{Obied
  et~al.}{2016}]{Obied2016}
Obied G.,  Zakamska N.~L.,  Wylezalek D.,   Liu G.,  2016, \mn@doi [Mon. Not.
  R. Astron. Soc.] {10.1093/mnras/stv2850}, 456, 2861

\bibitem[\protect\citeauthoryear{Pier \& Krolik}{Pier \&
  Krolik}{1992}]{Pier1992}
Pier E.~A.,  Krolik J.~H.,  1992, \mn@doi [Astrophys. J.] {10.1086/172042},
  401, 99

\bibitem[\protect\citeauthoryear{Pier \& Krolik}{Pier \&
  Krolik}{1993}]{Pier1993}
Pier E.~A.,  Krolik J.~H.,  1993, \mn@doi [Astrophys. J.] {10.1086/173427},
  418, 673

\bibitem[\protect\citeauthoryear{Pjanka, Greene, Seth, Braatz, Henkel, Lo  \&
  Laesker}{Pjanka et~al.}{2017}]{Pjanka2017}
Pjanka P.,  Greene J.~E.,  Seth A.~C.,  Braatz J.~A.,  Henkel C.,  Lo F. K.~Y.,
    Laesker R.,  2017, \mn@doi [The Astrophysical Journal, Volume 844, Issue 2,
  article id. 165, 20 pp. (2017).] {10.3847/1538-4357/aa7c18}, 844

\bibitem[\protect\citeauthoryear{Polletta et~al.,}{Polletta
  et~al.}{2007}]{Polletta2007}
Polletta M.,  et~al., 2007, \mn@doi [The Astrophysical Journal, Volume 663,
  Issue 1, pp. 81-102.] {10.1086/518113}, 663, 81

\bibitem[\protect\citeauthoryear{{Ramos Almeida} et~al.,}{{Ramos Almeida}
  et~al.}{2011}]{RamosAlmeida2011}
{Ramos Almeida} C.,  et~al., 2011, \mn@doi [Astrophys. J.]
  {10.1088/0004-637X/731/2/92}, 731, 92

\bibitem[\protect\citeauthoryear{Reyes et~al.,}{Reyes et~al.}{2008}]{Reyes2008}
Reyes R.,  et~al., 2008, \mn@doi [The Astronomical Journal, Volume 136, Issue
  6, pp. 2373-2390 (2008).] {10.1088/0004-6256/136/6/2373}, 136, 2373

\bibitem[\protect\citeauthoryear{Rose, Elvis, Crenshaw  \& Glidden}{Rose
  et~al.}{2015}]{rose2015}
Rose M.,  Elvis M.,  Crenshaw M.,   Glidden A.,  2015, \mn@doi [Monthly Notices
  of the Royal Astronomical Society: Letters, Volume 451, Issue 1, p.L11-L15]
  {10.1093/mnrasl/slv056}, 451, L11

\bibitem[\protect\citeauthoryear{Schmitt, Donley, Antonucci, Hutchings, Kinney
  \& Pringle}{Schmitt et~al.}{2003}]{Schmitt2003}
Schmitt H.~R.,  Donley J.~L.,  Antonucci R. R.~J.,  Hutchings J.~B.,  Kinney
  A.~L.,   Pringle J.~E.,  2003, \mn@doi [The Astrophysical Journal, Volume
  597, Issue 2, pp. 768-779.] {10.1086/381224}, 597, 768

\bibitem[\protect\citeauthoryear{Shlosman, Frank  \& Begelman}{Shlosman
  et~al.}{1989}]{Shlosman1989}
Shlosman I.,  Frank J.,   Begelman M.~C.,  1989, \mn@doi [Nature]
  {10.1038/338045a0}, 338, 45

\bibitem[\protect\citeauthoryear{Simpson}{Simpson}{2005}]{Simpson2005}
Simpson C.,  2005, \mn@doi [Monthly Notices of the Royal Astronomical Society,
  Volume 360, Issue 2, pp. 565-572.] {10.1111/j.1365-2966.2005.09043.x}, 360,
  565

\bibitem[\protect\citeauthoryear{Smee et~al.,}{Smee et~al.}{2012}]{Smee2012}
Smee S.,  et~al., 2012, \mn@doi [The Astronomical Journal, Volume 146, Issue 2,
  article id. 32, 40 pp. (2013).] {10.1088/0004-6256/146/2/32}, 146

\bibitem[\protect\citeauthoryear{Stalevski, Fritz, Baes, Nakos  \&
  Popovi{\'{c}}}{Stalevski et~al.}{2012}]{Stalevski2012}
Stalevski M.,  Fritz J.,  Baes M.,  Nakos T.,   Popovi{\'{c}} L.~{\v{C}}.,
  2012, \mn@doi [Mon. Not. R. Astron. Soc.] {10.1111/j.1365-2966.2011.19775.x},
  420, 2756

\bibitem[\protect\citeauthoryear{Treister, Krolik  \& Dullemond}{Treister
  et~al.}{2008}]{Treister2008}
Treister E.,  Krolik J.~H.,   Dullemond C.,  2008, \mn@doi [Astrophys. J.]
  {10.1086/586698}, 679, 140

\bibitem[\protect\citeauthoryear{Tristram, Burtscher, Jaffe, Meisenheimer,
  H{\"{o}}nig, Kishimoto, Schartmann  \& Weigelt}{Tristram
  et~al.}{2014}]{Tristram2014}
Tristram K. R.~W.,  Burtscher L.,  Jaffe W.,  Meisenheimer K.,  H{\"{o}}nig
  S.~F.,  Kishimoto M.,  Schartmann M.,   Weigelt G.,  2014, \mn@doi [Astron.
  Astrophys.] {10.1051/0004-6361/201322698}, 563, A82

\bibitem[\protect\citeauthoryear{{Veilleux} \& {Osterbrock}}{{Veilleux} \&
  {Osterbrock}}{1987}]{Veilleux1987}
{Veilleux} S.,  {Osterbrock} D.~E.,  1987, \mn@doi [\apjs] {10.1086/191166},
  \href {http://adsabs.harvard.edu/abs/1987ApJS...63..295V} {63, 295}

\bibitem[\protect\citeauthoryear{{Wright} et~al.,}{{Wright}
  et~al.}{2010}]{Wright2010}
{Wright} E.~L.,  et~al., 2010, \mn@doi [\aj] {10.1088/0004-6256/140/6/1868},
  \href {http://adsabs.harvard.edu/abs/2010AJ....140.1868W} {140, 1868}

\bibitem[\protect\citeauthoryear{{Wylezalek}, {Zakamska}, {Greene}, {Riffel},
  {Drory}, {Andrews}, {Merloni}  \& {Thomas}}{{Wylezalek}
  et~al.}{2018}]{Wylezalek2018}
{Wylezalek} D.,  {Zakamska} N.~L.,  {Greene} J.~E.,  {Riffel} R.~A.,  {Drory}
  N.,  {Andrews} B.~H.,  {Merloni} A.,   {Thomas} D.,  2018, \mn@doi [\mnras]
  {10.1093/mnras/stx2784}, \href
  {http://adsabs.harvard.edu/abs/2018MNRAS.474.1499W} {474, 1499}

\bibitem[\protect\citeauthoryear{Yan et~al.,}{Yan et~al.}{2015}]{Yan2015}
Yan R.,  et~al., 2015, \mn@doi [The Astronomical Journal, Volume 151, Issue 1,
  article id. 8, 18 pp. (2016).] {10.3847/0004-6256/151/1/8}, 151

\bibitem[\protect\citeauthoryear{Yan et~al.,}{Yan et~al.}{2016}]{Yan2016}
Yan R.,  et~al., 2016, \mn@doi [The Astronomical Journal, Volume 152, Issue 6,
  article id. 197, 32 pp. (2016).] {10.3847/0004-6256/152/6/197}, 152

\bibitem[\protect\citeauthoryear{{Zakamska} et~al.,}{{Zakamska}
  et~al.}{2005}]{Zakamska2005}
{Zakamska} N.~L.,  et~al., 2005, \mn@doi [\aj] {10.1086/427543}, \href
  {http://adsabs.harvard.edu/abs/2005AJ....129.1212Z} {129, 1212}

\bibitem[\protect\citeauthoryear{Zheng et~al.,}{Zheng et~al.}{2014}]{Zheng2014}
Zheng Z.,  et~al., 2014, \mn@doi [The Astrophysical Journal, Volume 800, Issue
  2, article id. 120, 26 pp. (2015).] {10.1088/0004-637X/800/2/120}, 800

\makeatother
\end{thebibliography}

\end{document}